\documentclass[a4paper,fleqn]{cas-dc}
\usepackage{caption}
\hypersetup{draft} 

\usepackage[numbers]{natbib}
\usepackage{amsmath,amssymb,amsfonts}
\usepackage{IEEEtrantools}
\usepackage{booktabs}
\usepackage{multirow}
\usepackage[flushleft]{threeparttable}
\usepackage{graphicx}
\usepackage{subcaption}
\usepackage{algorithm}
\usepackage{algpseudocode}
\usepackage{tikz}
\usetikzlibrary{arrows.meta, patterns, decorations.pathreplacing}
\usepackage[table]{xcolor}
\newtheorem{remark}{Remark}

\def\tsc#1{\csdef{#1}{\textsc{\lowercase{#1}}\xspace}}
\tsc{WGM}
\tsc{QE}
\tsc{EP}
\tsc{PMS}
\tsc{BEC}
\tsc{DE}

\begin{document}
\let\WriteBookmarks\relax
\def\floatpagepagefraction{1}
\def\textpagefraction{.001}

\shortauthors{A. Omidi et~al.}
\shorttitle{Constraint-Aware Reinforcement Learning Controller for BESS in PV Smoothing}
\title [mode = title]{Battery Energy Storage Sizing in PV Ramp-Rate Smoothing using Constraint-Aware Reinforcement Learning-Based Controller}

\author[1]{Arash Omidi}[type=editor,
                        auid=000,bioid=1]
\cormark[1]
\ead{arash.omidi@uvm.edu}
\credit{Writing - original draft, Conceptualization, Methodology, Software,
Formal analysis, Investigation, Validation, Visualization}

\affiliation[1]{organization={Department of Electrical and Biomedical Engineering, University of Vermont},
                city={Burlington},
                state={Vermont},
                country={USA}}

\author[1]{Tanmay Mishra}
\credit{Writing - review \& editing, Methodology, Visualization, Formal analysis}

\author[1]{Mads R. Almassalkhi}
\credit{Writing - review \& editing, Conceptualization, Supervision,
Project administration}

\cortext[cor1]{Corresponding author}


\ExplSyntaxOn
\RenewDocumentCommand \printorcid { }
{
  \seq_if_empty:NF \g_stm_orcid_seq
  {
    \group_begin:
      \tex_let:D \thefootnote \relax \footnotetext
      { \raggedright \textsc{orcid}(s):\c_space_token
        \seq_use:Nn \g_stm_orcid_seq { ;~ } }
    \group_end:
  }
}
\ExplSyntaxOff

\begin{abstract}
This paper presents a constraint-aware reinforcement learning (CARL) controller for battery energy storage system (BESS) sizing and photovoltaic (PV) ramp-rate smoothing. Rapid PV power fluctuations caused by cloud transitions can introduce operational and stability challenges in power systems. To mitigate these impacts, grid codes impose ramp-rate limits (RRL) on PV power injected to the grid. BESSs can support such grid codes by absorbing or supplying power during rapid variations. However, the extent to which BESSs can reduce PV-induced variability depends on (i) the BESS power and energy capacities; and (ii) the adopted controller. Thus, BESS sizing and controller design are coupled. Conventional moving-average control can increase battery usage due to filter-induced lag, while ramp-rate control is reactive and does not account for the long-term impact of actions on battery state of charge and future ramp events. Optimization-based methods can reduce BESS requirements by anticipating future PV variations, but their performance depends on accurate PV forecasts. To address these limitations, the CARL controller learns a causal state-feedback policy from historical PV data without requiring forecasts during deployment. Ramp-rate, converter-power, and state-of-charge constraints are incorporated into a causal feasibility interval that guarantees feasible actions by definition. A full-horizon input-convex neural network (ICNN)-based optimization is developed as a noncausal benchmark for the minimum achievable BESS requirements. Evaluation using three real-world PV datasets shows that, at a 10\%/min ramp-rate limit, CARL reduces peak BESS power by 8.7--10.4\% and energy capacity by 47.5--62.6\% relative to conventional ramp-rate control while maintaining the grid-codes compliance.
\end{abstract}





\maketitle

\section{Introduction}

The global transition toward low-carbon power systems has placed solar photovoltaic (PV) generation at the center of the energy mix because of its environmental and economic benefits, with the installed capacity expected to increase by approximately 3.7~TW over the period 2025 to 2030~\cite{IEA2025Renewables}. However, rapid fluctuations in solar irradiance, particularly due to cloud transition and meteorological conditions, can lead to abrupt changes in PV power output over short time intervals, which pose operational and stability challenges for power systems, especially isolated grids, microgrids, or grids without spatially dispersed PV installations~\cite{van2014empirical}. These effects can result in frequency deviation, voltage instability, and increased stress on dispatchable generation units tasked with balancing supply and demand mismatch~\cite{sukumar2018ramp,tran2018mitigation}. 

To mitigate these adverse effects, grid codes impose ramp-rate limits (RRL) on PV power output, specifying the maximum allowable change in output over a defined time interval (e.g.~\%/min)~\cite{gevorgian2013review}.  For example, reported RRL include $\pm10\%/\mathrm{min}$ imposed by the Puerto Rico Electric Power Authority (PREPA), $\pm1$--$2\mathrm{MW}/\mathrm{min}$ in Hawaii, $+10\%/\mathrm{min}$ in Germany, and $\pm100\mathrm{kW}/\mathrm{s}$ in Denmark~\cite{gevorgian2013review,energinet2015technical,zheng2017overivew}. 

To meet these requirements, two approaches are commonly used. One approach is active power curtailment, which limits the rate of increase in PV power output during rapid irradiance variations. However, it cannot mitigate downward ramps and results in energy losses due to curtailed generation~\cite{sangwongwanich2016cost}. An alternative strategy employs energy storage systems, which absorb excess power during upward ramps and inject power during downward ramps, thereby enabling mitigation of PV fluctuations in both directions~\cite{li2013battery,papageorgiou2024adaptive}.

Among the storage technologies employed for ramp-rate mitigation~\cite{bullich2020review}, lithium-ion battery energy storage systems (BESS) are widely adopted due to high energy density, high efficiency, and declining cost~\cite{beltran2019levelized}. However, BESS effectiveness is governed by its power and energy capacities relative to PV variability, making sizing a critical design consideration. An undersized system fails to compensate worst-case ramps, causing RRL violations, while an oversized one inflates capital costs and operational losses. The required power and energy capacities are jointly determined by PV variability and the BESS dispatch imposed by the controller. Therefore, sizing cannot be treated as an independent step, but must be co-optimized with the control strategy to achieve both technical compliance and economic viability.

PV smoothing methods determine the BESS response to PV power fluctuations, and are generally categorized as filter-based, ramp-rate-based, optimization-based, and machine learning-based approaches~\cite{sukumar2018ramp}. Each method produces a distinct BESS dispatch profile and therefore influences the required power and energy capacities. Filter-based methods, such as the moving average control (MAC), are simple to implement but introduce a phase lag between the PV power and the smoothed reference~\cite{kini2019mitigation,cano2022comparative,koiwa2017analysis}. This lag must be compensated by the BESS, which increases the energy throughput and potentially the required BESS size~\cite{marcos2014control}. Conventional ramp-rate control (RRC) activates the BESS only when PV ramp rate exceeds the limit, thereby reducing unnecessary cycling compared with filter-based methods~\cite{sukumar2018ramp}. However, because RRC reacts only after a ramp-rate violation occurs, it may require a large instantaneous power response, and additionally, without explicit state-of-charge (SoC) management, repeated charging or discharging can drive the battery toward its operating limits~\cite{huo2020novel}. Moreover, MAC and RRC also do not explicitly optimize BESS power and energy capacities subject to operational constraints. Optimization-based methods, such as model predictive control, can jointly account for ramp-rate compliance, SoC regulation, and BESS sizing while anticipating future PV variations, thereby reducing required battery capacity~\cite{abdelghany2024integration,chen2018forecasting}. Their performance, however, depends on accurate PV forecasts. Forecasting methods based on satellite imagery, all-sky cameras, or on-site sensors may not be widely available, and forecast errors generally increase with prediction horizon because of cloud-motion uncertainty~\cite{ryu2021battery,wen2020deep,urquhart2013sky,paletta2022eclipse}. These limitations motivate data-driven control strategies that can learn from historical PV trajectories to capture the sizing benefits of predictive methods while remaining causal and avoiding explicit forecasts during deployment.

Machine learning approaches such as reinforcement learning (RL) offer a compelling framework to address these limitations. Unlike conventional approaches, RL learns directly from experience, where an agent interacts with a simulated environment built from historical PV data and BESS, and iteratively refines its control policy to maximize the cumulative long-term reward~\cite{perera2021applications}. This learning paradigm enables the agent to learn effective control policies not only under stable operating conditions but also in the presence of inherent system uncertainty and variability. RL is particularly suitable for PV smoothing because each BESS action affects not only the current ramp-rate response, but also the future SoC trajectory, available power and energy headroom, and ultimately required BESS size. Despite this potential, most RL applications in PV-plus-storage systems target energy arbitrage or microgrid  scheduling~\cite{huang2020deep,arwa2020reinforcement}, while existing studies on PV ramp mitigation do not address BESS sizing. For example, RL has been used for battery-less PV smoothing through curtailment~\cite{chen2023robust} and for PV forecasting for intraday economic dispatch with fixed storage  capacity~\cite{cheng2023mitigating}. Consequently, there remains a research gap in the development of causal PV-smoothing policies that jointly reduce BESS power and energy requirements while maintaining RRL compliance and BESS constraints.

To address this gap, this work proposes a constraint-aware reinforcement learning (CARL) framework that restricts the agent to actions satisfying the ramp-rate, SoC, and power constraints while learning a causal policy that reduces the required BESS power and energy requirements without relying on PV forecasts. Compared with the existing literature on PV smoothing, our contribution differs in five important respects:
\begin{itemize}
    \item PV ramp-rate smoothing is formulated as an RL problem that explicitly targets reducing the required BESS power and energy capacities while maintaining grid-code compliance.

    \item The RRL is enforced by definition, rather than through reward penalties, by mapping the policy action into a causal feasibility interval defined by the ramp-rate, converter power, and SoC constraints, thereby guaranteeing feasible operation at every step.

    \item An iterative simulation-based BESS sizing procedure is developed to determine the required power and energy capacities for a given smoothing controller.
    
    \item A noncausal optimization benchmark is formulated using complete PV trajectory information to provide a lower bound on the required BESS power and energy capacities.

    \item The technical and economic performance of the proposed controller is evaluated against MAC, RRC, and the optimization-based benchmark. The controller is also validated on three real-world datasets with different PV power variability and capacity. 
   
\end{itemize}

The remainder of this paper is organized as follows. Section~\ref{sec:System Model and Problem Formulation} presents the PV-BESS system model and operational constraints. Section~\ref{sec:sizing_algorithm} formulates the PV power smoothing problem and describes the BESS sizing framework. Section~\ref{sec: proposed RL} introduces the proposed CARL controller, including the a Markov decision process formulation, feasibility-constrained action mapping, reward function, and training procedure. Section~\ref{sec:Smoothing Strategies} describes the benchmark PV smoothing strategies. Section~\ref{sec:simulation and results} presents the simulation results and evaluates the proposed method against the benchmark controllers. Finally, Section~\ref{sec:conclusion} concludes the paper.

\section{System Model and Constraints}
\label{sec:System Model and Problem Formulation}
\begin{figure}
    \centering
   \includegraphics[width=0.9\columnwidth]{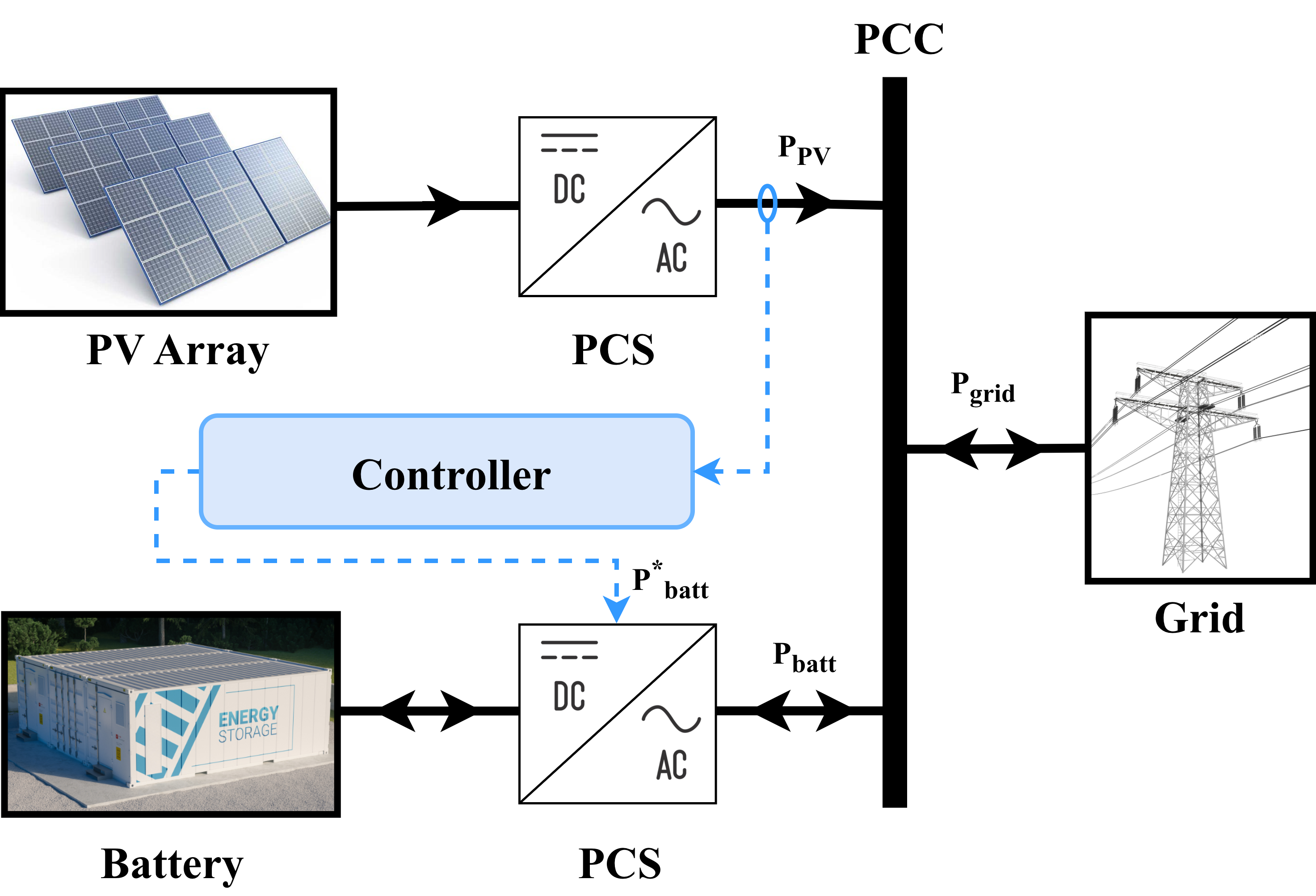}
    \caption{AC-coupled PV-BESS system configuration. The PV plant and BESS are interfaced to the PCC through their power converter systems (PCSs), with the controller dispatching the battery setpoint $P^{*}_{\mathrm{batt}}$ at each control step.}
    \label{fig:system_model}
\end{figure}

Fig.~\ref{fig:system_model} illustrates the configuration of the system under study, consisting of a grid-connected PV plant integrated with a BESS. The PV plant and BESS are connected to the grid at the point of common coupling (PCC) through separate power converter systems. The net active power delivered to the grid ($P_{\mathrm{grid}}[k]$) at the PCC is given by
{\small\begin{equation}
    \label{powerflow}
    P_{\mathrm{grid}}[k] = P_{\mathrm{PV}}[k] - P_{\mathrm{batt}}[k], \qquad    k \in \mathcal{K} = \{0,1,\dots,N-1\}
\end{equation}}
where $k$ denotes the discrete-time index and $N$ is the total number of time steps in one day. In this study, $N=1440$, representing 24 hours of data sampled at one-minute intervals. $P_{\mathrm{PV}}[k]$ denotes the instantaneous raw PV power delivered by the PV plant to the PCC, and $P_{\mathrm{batt}}[k]$ represents the power exchanged between BESS and the PCC at time step $k$. The sign convention adopted here defines $P_{\mathrm{batt}}[k]>0$ as charging and $P_{\mathrm{batt}}[k]<0$ as discharging. The controller, shown in Fig.~\ref{fig:system_model}, operates the BESS according to the specified control objective by selecting the BESS power setpoint, $P^\star_{\mathrm{batt}}[k]$, at each time step $k$. Since the BESS power converter control loops operate at a much faster timescale than the plant controller, setpoint tracking is assumed to be ideal at the one-minute resolution of this study, i.e., $P_{\mathrm{batt}}[k]=P^\star_{\mathrm{batt}}[k]$. In this study, the control objective is to smooth the PV power delivered to the grid such that $P_{\mathrm{grid}}[k]$ satisfies the RRL defined later in Section~\ref{sec: Ramp-Rate constraint section}. The feasibility of a smoothing strategy depends on the power and energy limits of the BESS, which is modeled next.

\subsection{Battery Energy Storage System Model}
\label{sec:BESS model}
The BESS is represented by an energy reservoir model, which is widely used for battery operation in power systems studies. The discretized SoC dynamics, obtained using zero-order-hold sampling with time step  $\Delta t$, are given by

\begin{subequations}
{\footnotesize
\begin{align}
    E_{\mathrm{batt}}[k+1] =& E_{\mathrm{batt}}[k] + \tfrac{\Delta t}{60\times C_{\mathrm{batt}}} \left(\eta_{\mathrm{ch}}[k] P_{\mathrm{batt}}^{\mathrm{ch}}[k] - \frac{P_{\mathrm{batt}}^{\mathrm{dis}}[k]}{\eta_{\mathrm{dis}}[k]} \right)
    \label{Soc_model} \raisetag{1.5\baselineskip} \\
    &\eta_{\mathrm{ch}}[k] = f_{\mathrm{ch}} (P_{\mathrm{batt}}^{\mathrm{ch}}[k])\\
    &\eta_{\mathrm{dis}}[k] = f_{\mathrm{dis}} (P_{\mathrm{batt}}^{\mathrm{dis}}[k])\\
    &\underline{E}_{\mathrm{batt}}\leq E_{\mathrm{batt}}[k] \leq \overline{E}_{\mathrm{batt}},\\
    &0 \leq P_{\mathrm{batt}}^{\mathrm{ch}}[k], P_{\mathrm{batt}}^{\mathrm{dis}}[k] \leq {P}^\mathrm{max}_{\mathrm{batt}}, \label{battery_constraint}\\
    &P_{\mathrm{batt}}^{\mathrm{ch}}[k] P_{\mathrm{batt}}^{\mathrm{dis}}[k] =0, \\
    &P_{\mathrm{batt}}[k] = P_{\mathrm{batt}}^{\mathrm{ch}}[k] - P_{\mathrm{batt}}^{\mathrm{dis}}[k], \label{battery_power}
\end{align}} \label{ERM}
\end{subequations}
\par\noindent
where $E_{\mathrm{batt}}[k]$ is the SoC at time step $k$, and $P_{\mathrm{batt}}^{\mathrm{ch}}[k]$ and $P_{\mathrm{batt}}^{\mathrm{dis}}[k]$ denote the AC-side BESS charging and discharging powers, respectively, for all time steps $k \in \mathcal{K}$. $C_{\mathrm{batt}}$ is the nominal capacity of the battery in kWh, and the charging and discharging efficiencies at time $k$ are given by $\eta_\text{ch}[k], \eta_\text{dis}[k] \in (0,1]$, respectively. The majority of PV smoothing studies assume constant charging and discharging efficiencies~\cite{marcos2014control,tahir2022optimal,amorim2024sizing}. In practice, however, the efficiencies are nonlinear functions of BESS power, showing a sharp decline under part-load operating conditions ($<0.3$ p.u.). Since the BESS operates across a wide range of loading conditions throughout the day, a fixed efficiency value fails to represent the true conversion losses over the full operating range, resulting in a mismatch between the predicted and actual SoC~\cite{omidi2025towards}. This error propagates directly into the sizing procedure. As a result, the required BESS capacity may be over- or underestimated depending on the chosen constant efficiency and the PV power profile.

To avoid this bias, the nonlinear efficiency is modeled using the measured efficiency data from a utility-scale inverter reported in~\cite{Faranda}. At each step, the charging and discharging efficiencies are obtained from a one-dimensional look-up table indexed by the per-unit converter loading, with linear interpolation between the measured breakpoints.
Equation \eqref{ERM} fully characterizes the BESS model used in this work. The SoC dynamics, nonlinear efficiency, and power limits jointly determine the range of compensation the BESS can provide at each time step. The extent to which this range can satisfy the grid-imposed ramp-rate requirement motivates the feasibility analysis presented in the following section.

\subsection{Ramp-Rate Limit and Feasibility Interval}
\label{sec: Ramp-Rate constraint section}

The per-step ramp rate of the plant output at the PCC is defined as the normalized rate of change of the $P_\mathrm{grid}$ over each control interval,
\par \noindent
{\small
\begin{equation}
     \mathcal{R}[k]=\frac{(P_\mathrm{grid}[k] - P_\mathrm{grid}[k-1])\times 100}{P_\mathrm{pv}^\mathrm{rated} \Delta t} \qquad |\mathcal{R}[k]|\leq\mathcal{R}_\mathrm{max},\label{ramp_rate_equation}
\end{equation}}
where $\mathcal{R}_\mathrm{max}$ is the maximum admissible ramp rate, typically $10\%/\min$ for utility-scale plants~\cite{marcos2014control}. The control interval $\Delta t$ is expressed in minutes, and $P_{\mathrm{pv}}^{\mathrm{rated}}$ denotes the rated PV power. Equivalently, the admissible plant output at step $k$ is bounded within the closed interval,

{\footnotesize
\begin{align}
   &P_\mathrm{grid}[k] \;\in\; \big[\,P_\mathrm{grid}[k-1]- \Delta P_\mathrm{max},\; P_\mathrm{grid}[k-1] + \Delta P_\mathrm{max}\,\big],\label{grid power constraints}\\
    &\Delta P_\mathrm{max} := \frac{\mathcal{R}_\mathrm{max}\,P_\mathrm{pv}^\mathrm{rated}\,\Delta t}{100}.
\end{align}}
\par \noindent
Substituting the AC-side power balance~\eqref{powerflow} into~\eqref{grid power constraints}, the admissible BESS power setpoint at step $k$ is bounded by the lower and upper RRL as
\begin{subequations}
    \begin{align}
        \underline{P}^{\mathrm{ramp}}_\mathrm{batt}[k] = P_\mathrm{PV}[k] - P_\mathrm{grid}[k-1] - \Delta P_\mathrm{max}, \\
        \overline{P}^{\mathrm{ramp}}_\mathrm{batt}[k] = P_\mathrm{PV}[k]- P_\mathrm{grid}[k-1] + \Delta P_\mathrm{max},
    \end{align}
    \label{battery ramp rate constraint}
\end{subequations}
where $\underline{P}^{\mathrm{ramp}}_\mathrm{batt}[k]$ and 
$\overline{P}^{\mathrm{ramp}}_\mathrm{batt}[k]$ represent the minimum and maximum  admissible BESS power setpoints at step $k$, respectively, ensuring that~\eqref{ramp_rate_equation} is satisfied for all $k$.

A second operational constraint arises from the available energy capacity of the BESS. The dynamics of the BESS SoC over each control interval are determined by the charging and discharging powers and the associated conversion efficiencies, as described by~\eqref{Soc_model}. Since the conversion efficiencies depend on BESS power, an exact inversion of~\eqref{Soc_model} leads to an implicit bound for BESS power. To derive an explicit and guaranteed-feasible bound, the efficiencies are conservatively represented by their worst-case values over the admissible power range. Enforcing $E_{\mathrm{batt}}[k]\in [\underline{E}_{\mathrm{batt}},\overline{E}_{\mathrm{batt}}]$
yields the maximum charging and discharging powers admissible based on the present SoC,
\begin{subequations}
\begin{align}
    \overline{P}^{\mathrm{soc}}_\mathrm{batt}[k] &= \frac{\big(\overline{E}_{\mathrm{batt}}-E_{\mathrm{batt}}[k]\big)\,60 \times C_\mathrm{batt}}
    {\eta_\mathrm{ch}^\mathrm{max} \Delta t}, \label{Psoc_up_wc}\\[2pt]
    \underline{P}^{\mathrm{soc}}_\mathrm{batt}[k] &= -\,\frac{\big(E_{\mathrm{batt}}[k]-\underline{E}_{\mathrm{batt}}\big)\, 60\times\eta_\mathrm{dis}^\mathrm{min}C_\mathrm{batt}}{\Delta t},
    \label{Psoc_dn_wc}
\end{align}
\label{Psoc_wc}
\end{subequations}
where $\eta_\mathrm{ch}^{\max}=\max_{P}\eta_\mathrm{ch}(P_\mathrm{batt})$ and
$\eta_\mathrm{dis}^{\min}=\min_{P}\eta_\mathrm{dis}(P_\mathrm{batt})$ over $(0,P_\mathrm{batt}^\mathrm{max}]$. This conservative representation yields the explicit bounds in~\eqref{Psoc_wc} while ensuring satisfaction of the SoC constraints. With appropriate BESS sizing and control strategy, this bound is rarely restrictive, and the BESS operates near the midpoint of the SoC range.
\begin{figure}
    \centering
    \includegraphics[width=\columnwidth]{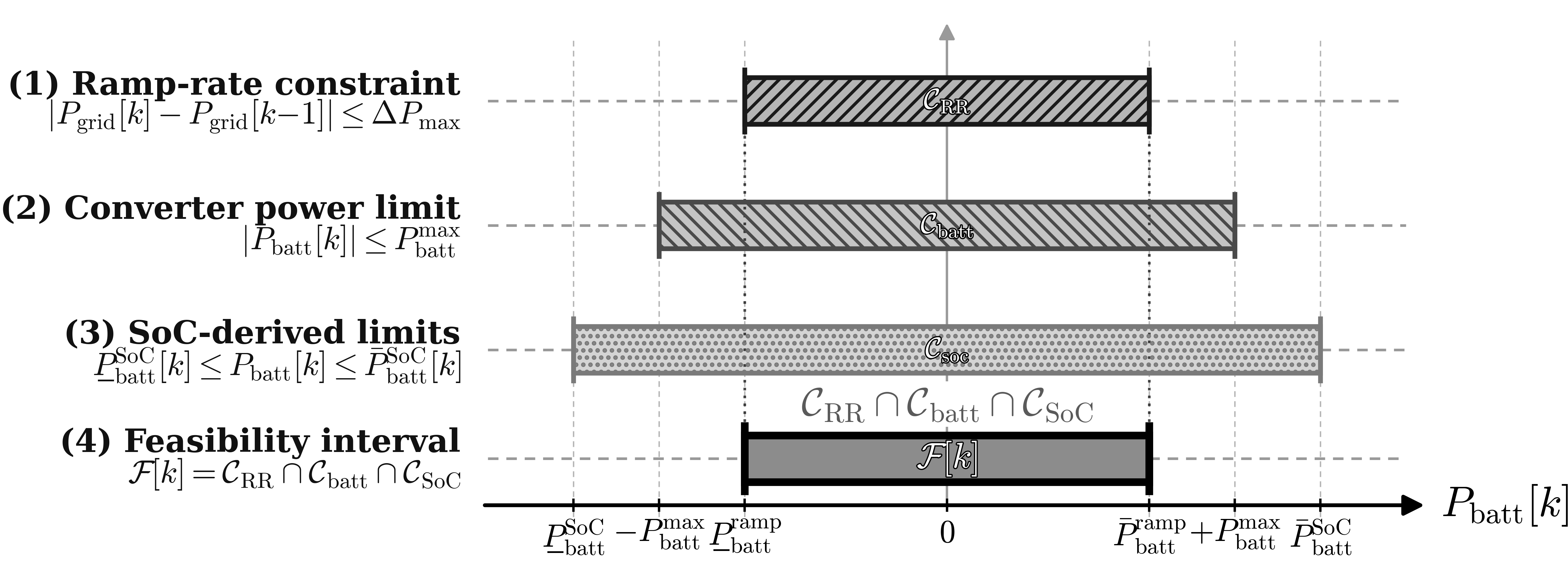}
    \caption{Construction of the feasibility interval $\mathcal{F}[k]$ as the intersection of three constraint sets: the ramp-rate constraint $\mathcal{C}_{\mathrm{RR}}$, the BESS power limit $\mathcal{C}_{\mathrm{batt}}$, and the SoC-derived limit $\mathcal{C}_{\mathrm{SoC}}$}
    \label{fig:feasibility}
\end{figure}
As shown in Fig.~\ref{fig:feasibility}, intersecting~\eqref{battery ramp rate constraint} with the converter maximum power bound~\eqref{battery_constraint} and the SoC admissible band~\eqref{Psoc_wc} yields the feasibility interval for the BESS power:
\begin{equation}
    P_\mathrm{batt}[k] \;\in\; \mathcal{F}[k] \;:=\; \big[\,\underline{P}_\mathrm{batt}[k],\, \overline{P}_\mathrm{batt}[k]\,\big],
    \label{feasibility_interval}
\end{equation}
with endpoints
\begin{subequations}{
\small
\begin{align}
\underline{P}_\mathrm{batt}[k] = \max\!\Big\{\, \underline{P}^{\mathrm{ramp}}_\mathrm{batt}[k],\;\; -P_\mathrm{batt}^\mathrm{max},\;\; \underline{P}_\mathrm{batt}^\mathrm{soc}[k]\Big\}, \\
\overline{P}_\mathrm{batt}[k]  = \min\!\Big\{\, \overline{P}^{\mathrm{ramp}}_\mathrm{batt}[k],\;\; +P_\mathrm{batt}^\mathrm{max},\;\; \overline{P}_\mathrm{batt}^\mathrm{soc}[k]\Big\}.
\end{align}}
\label{Feas_interval}
\end{subequations}
The feasibility interval $\mathcal{F}[k]$ is by definition non-empty whenever the BESS is sized large enough to absorb the worst-case ramp event observed at the site. In the rare cases where $\underline{P}_\mathrm{batt}[k] > \overline{P}_\mathrm{batt}[k]$, the BESS cannot fully satisfy the imposed constraints, indicating that the selected power and energy limits are insufficient for the corresponding operating condition. The interval $\mathcal{F}[k]$ has two structural properties that are central to the rest of the paper as follows:
\begin{itemize}
    \item It depends only on 
the present and previous measurements, i.e., $P_\mathrm{PV}[k]$, $P_\mathrm{grid}[k-1]$, and $\mathrm{E}[k]$, ensuring $\mathcal{F}[k]$ is a causal set.
\item Any setpoint chosen inside $\mathcal{F}[k]$ guarantees ramp-rate compliance at step $k$, since the constraint~\eqref{ramp_rate_equation} is enforced by definition.
\end{itemize}

Once $\mathcal{F}[k]$ is constructed, ramp-rate compliance is guaranteed for any admissible controller. The only remaining freedom is which BESS power setpoint within $\mathcal{F}[k]$ to select at each step, a sequential decision that influences the future feasibility intervals and the required BESS power and energy. Accordingly, the following section formulates the PV power smoothing problem and introduces the BESS sizing framework used to quantify BESS requirements.

\section{PV Power Smoothing and BESS Sizing Framework}
\label{sec:sizing_algorithm}
The PV power smoothing controller determines the BESS power setpoint, $P_\mathrm{batt}^{\star}[k]$, at each time step $k$ to regulate the power delivered to the grid, $P_\mathrm{grid}[k]$, while satisfying the RRL and BESS constraints discussed in Section~\ref{sec:System Model and Problem Formulation}. In general, a smoothing controller is represented by
\begin{equation}
P_\mathrm{batt}^{\star}[k]= f_c\left(\mathbf{x}[k]\right), \qquad f_c\left(\mathbf{x}[k]\right) \in \mathcal{F}[k],
\label{general_smoothing_controller}
\end{equation}
where $f_c(\cdot)$ denotes the smoothing controller, and $\mathbf{x}[k]$ contains the input variables available to the controller, which may include current and historical PV power, previous grid power, and battery SoC. The smoothing strategies differ primarily in how $f_c(\cdot)$ selects $P_\mathrm{batt}^{\star}[k]$. Since the resulting BESS dispatch directly determines the required BESS power and energy capacities, the battery size is inherently dependent on the adopted smoothing strategy. 

Existing BESS sizing approaches for PV smoothing generally determine the required battery size based on either worst-case ramp events or long-term simulation using representative PV power profiles~\cite{amorim2024sizing,lappalainen2022sizing}. Worst-case-event sizing characterizes BESS requirements from individual extreme ramps and may not capture the cumulative effect of the smoothing controller on battery power and SoC. In contrast, simulation-based sizing evaluates the controller over PV trajectories and therefore accounts for the temporal utilization of the BESS. Accordingly, this study develops a simulation-based BESS sizing algorithm, detailed in Algorithm~\ref{alg:sizing}.

The sizing procedure can be initialized with sufficiently large BESS power and energy ratings to accommodate the PV ramp-rate requirements. For a candidate BESS with power rating $P_{\mathrm{batt}}^{\max}$ and nominal energy capacity $C_{\mathrm{batt}}$, the corresponding smoothing controller is evaluated for each day $d$ in the sizing dataset $\mathcal{D}$. The required power rating is the maximum absolute BESS power across all time steps and days and the required energy capacity is obtained from the maximum SoC deviation from its reference value, $E_{\mathrm{batt}}^{\mathrm{ref}}$, as defined by
\begin{subequations} 
\label{sizing} 
\begin{align} 
P_{\mathrm{batt}}^{\mathrm{req}} &= \max_{d\in\mathcal{D}} \max_{k} \left| P_{\mathrm{batt}}[k;d] \right|, \label{sizing_power} \\ 
C_{\mathrm{batt}}^{\mathrm{req}} &= 2C_{\mathrm{batt}} \displaystyle\max_{d\in\mathcal{D}}\max_{k} \left| E_{\mathrm{batt}}[k;d] - E_{\mathrm{batt}}^{\mathrm{ref}} \right|. \label{sizing_energy} \end{align}
\end{subequations}
Since insufficient ratings may lead to ramp-rate violations, the procedure checks RRL compliance and increases the power and energy ratings when violations occur. The resulting ratings are then adjusted to satisfy the C-rating requirements.

\begin{remark}
    Because the BESS conversion efficiency is power-dependent, changes in the power rating affect the estimated charging and discharging efficiencies and, consequently, the battery energy trajectory and required energy capacity. Therefore, the controller is re-evaluated iteratively until the BESS power rating converges.
\end{remark}

This procedure provides a consistent basis for comparing the BESS requirements of the considered smoothing strategies. Motivated by the dependence of these requirements on the adopted control policy, the following section introduces the proposed constraint-aware reinforcement learning controller (CARL) for PV ramp-rate smoothing.
 \begin{algorithm}[t] 
 \caption{BESS Sizing} \label{alg:sizing}
 \small
 \begin{algorithmic}[1]
\Require Smoothing controller $f_c$, dataset $\mathcal{D}$, initial BESS ratings $P_{\mathrm{batt}}^{\max,(0)}$ and $C_{\mathrm{batt}}^{(0)}$, $C_{\mathrm{rate}}$, power and energy increase factors $\delta_P$ and $\delta_C$, convergence tolerances $\epsilon_P$, and maximum number of iterations $I_{\max}$

 \Ensure Sized BESS ratings $P_{\mathrm{batt}}^{\max}$ and $C_{\mathrm{batt}}$ 
 \State Initialize iteration index $i \leftarrow 0$ 
 \Repeat 
 \State Set the BESS ratings to $P_{\mathrm{batt}}^{\max,(i)}$ and $C_{\mathrm{batt}}^{(i)}$ 
 \State Run controller $f_c$ over all $d\in\mathcal{D}$ 
 \State Record $P_{\mathrm{batt}}^{(i)}[k;d]$, $E_{\mathrm{batt}}^{(i)}[k;d]$, and  $\mathcal{R}^{(i)}[k;d]$
 \State Compute $P_{\mathrm{batt}}^{\mathrm{req}} \text{ and } C_{\mathrm{batt}}^{\mathrm{req}} \leftarrow \eqref{sizing_power} \text{ and } \eqref{sizing_energy}$
 \State $V_{\mathrm{RR}} \leftarrow
    \exists\, d\in\mathcal{D},\,k\in\mathcal{K}:
    |\mathcal{R}^{(i)}[k;d]|>\mathcal{R}_\mathrm{max}$
\If{$V_{\mathrm{RR}}$}
    \State \textbf{if} $P_{\mathrm{batt}}^{\mathrm{req}} \geq P_{\mathrm{batt}}^{\max,(i)}$ \textbf{then} $P_{\mathrm{batt}}^{\mathrm{req}} \leftarrow (1+\delta_P)P_{\mathrm{batt}}^{\max,(i)}$
    \State \textbf{if} $C_{\mathrm{batt}}^{\mathrm{req}} \geq C_{\mathrm{batt}}^{(i)}$ \textbf{then} $C_{\mathrm{batt}}^{\mathrm{req}} \leftarrow (1+\delta_C)C_{\mathrm{batt}}^{(i)}$
\EndIf
 \State $\Delta_P \leftarrow \left| P_{\mathrm{batt}}^{\mathrm{req}} - P_{\mathrm{batt}}^{\max,(i)} \right|$ 
 \State $C_{\mathrm{batt}}^{\mathrm{req}} \leftarrow
\max\left\{
C_{\mathrm{batt}}^{\mathrm{req}},
P_{\mathrm{batt}}^{\mathrm{req}}/C_{\mathrm{rate}}^{\max}
\right\}$
 \State Update $P_{\mathrm{batt}}^{\max,(i+1)} \leftarrow P_{\mathrm{batt}}^{\mathrm{req}} $ and $C^{(i+1)}_{\mathrm{batt}} \leftarrow C_{\mathrm{batt}}^{\mathrm{req}} $
 \State $i \leftarrow i+1$ 
 \Until{ $\Delta_P \leq \epsilon_P$ or $i = I_{\max}$\textbf{} }
 \State \Return $P_{\mathrm{batt}}^{\max,(i)}$ and $C_{\mathrm{batt}}^{(i)}$
 \end{algorithmic} 
 \end{algorithm}

\section{Proposed Reinforcement-Learning Controller}
\label{sec: proposed RL}
Reinforcement learning (RL) is a sequential decision-making framework in which an agent interacts with an environment, observes the system states, selects an action according to a policy, and receives a reward. Through repeated interactions, the agent learns a policy that maximizes the expected cumulative reward~\cite{sutton1998reinforcement}. Unlike reactive controllers, RL can account for the long-term effects of current decisions on future system states and performance. This is particularly relevant for PV smoothing with a BESS, where each decision not only affects the current grid ramp rate, but also changes the battery energy state and the available control flexibility at subsequent time steps. Therefore, RL provides a framework for learning long-term control behavior from historical PV trajectories while retaining causal operation at deployment. The PV smoothing problem therefore can be represented as an RL problem, where the sequential interaction is modeled as a Markov decision process (MDP). 

\subsection{PV Smoothing as a Markov Decision Process}
\label{sec:MDP}
In this work, the PV smoothing problem is formulated as a finite-horizon MDP with state-dependent hard constraints on the physical BESS action. Rather than incorporating violations of the ramp rate, BESS power, and SoC limits as penalties in the reward function, these operational constraints are enforced directly through the feasibility interval derived in Section~\ref{sec: Ramp-Rate constraint section}. Therefore, the RL policy learns to select among feasible BESS operating points while optimizing the long-horizon BESS power and energy requirement. The finite-horizon MDP is defined as \begin{equation}
\mathcal{M} = \left( \mathcal{S}, \mathcal{A}, \mathcal{P}, r, \gamma \right), \label{eq:mdp_definition} 
\end{equation}
where $\mathcal{S}$, $\mathcal{A}$, $\mathcal{P}$, $r$, $\gamma$ denote the state space, action space, transition dynamics, reward, and discount factor, respectively.
In the present problem, the controller acts as agent and the measured PV profile with the BESS are the environment. At each time step $k$, the agent observes the state $s_k \in \mathcal{S}$, selects an action $a_k \in \mathcal{A}$ according to a policy $\pi_\theta: \mathcal{S} \to \mathcal{A}$, and receives a reward $r_k = r(s_k, a_k)$. The BESS dynamics and next PV measurements determine the next state $s_{k+1}$. In addition, the long-term performance is quantified by the return $G_k = \sum_{i=k}^{N-1} \gamma^{\,i-k} r(s_i, a_i)$, defined as the discounted sum of rewards, wherein $\gamma \in [0,1)$ is the discount factor that balances immediate and future rewards and $N$ denotes the episode horizon. The objective is to determine the optimal policy $\pi_\theta$, parameterized by $\theta$, that maximizes the expected long-term return. To apply this framework to the PV smoothing task, the constituent elements of the decision process must be specified explicitly, as detailed in the following.

\subsubsection{State Space}
The state contains the information available to the agent about the current condition of the environment for selecting the control action. At time step $k$, $s_{\text{k}}$ is defined as
{\small
\begin{equation} 
s_k = \left( \boldsymbol{p}_{\mathrm{PV}}[k], P_{\mathrm{grid}}[k-1], E_{\mathrm{batt}}[k], P_{\mathrm{batt}}^{\mathrm{peak}}[k], \tau_k \right), 
\label{state}
\end{equation}
}where $\boldsymbol{p}_\mathrm{PV}[k]:= [P_\mathrm{PV}[k-H+1],\dots,P_\mathrm{PV}[k]]^T \in \mathbb{R}^{H}$ contains the most recent $H$ PV measurements, with $H$ treated as a hyperparameter. This history captures recent PV variations without using forecasts. $P_{\mathrm{grid}}[k-1]$ determines the admissible ramp at the current step, while $E_{\mathrm{batt}}[k]$ represents the present battery SoC. $P_{\mathrm{batt}}^{\mathrm{peak}}[k]$ is running peak BESS power, defined as
\begin{equation}
P_{\mathrm{batt}}^{\mathrm{peak}}[k] = \max_{0\leq j<k}\left|P_{\mathrm{batt}}[j]\right|,
\label{running_peak}
\end{equation}
and provides the agent with information regarding the largest BESS power requirement established until this time step. Finally, $\tau_k=\frac{k}{N-1}$ provides normalized temporal information within the daily horizon.

\subsubsection{Action and Feasibility Map } 
A conventional RL formulation may allow the agent to select BESS power over an unrestricted range and penalize violations of the ramp-rate, BESS power, and SoC constraints through the reward function. However, such penalty-based methods do not guarantee constraint satisfaction and allows agents to explore physically infeasible actions during training. In addition, the resulting level of constraint enforcement would depend on the relative tuning of the reward penalties. To avoid these challenges, the proposed constraint-aware reinforcement learning (CARL) framework incorporates operational constraints directly into the action space. At each time step $k$, the feasible BESS power interval $\mathcal{F}[k]$, derived in~\eqref{feasibility_interval}, restricts the control action to physically admissible operating points.
The agent produces a normalized action $a_k \in [-1,1]$, which is mapped to $\mathcal{F}[k]$ through

{\small
\begin{equation}
    P_\mathrm{batt}[k] \;=\; \underline{P}_\mathrm{batt}[k] \;+\; \tfrac{1}{2}\big(a_k + 1\big)\big(\overline{P}_\mathrm{batt}[k] - \underline{P}_\mathrm{batt}[k]\big), \label{action_feas}
\end{equation}}
where $a_k = -1$ and $a_k = +1$ select the lower and upper feasible bounds, respectively. This mapping guarantees feasible actions by definition, allowing the agent to focus on selecting the operating point that optimizes the long-term objective.

\subsubsection{Reward Function}
Since the feasibility map already enforces the hard operational constraints, the reward function is used to rank feasible actions according to the BESS sizing objective. Specifically, the reward is designed to discourage battery usage, reduce SoC deviation from the reference, and peak BESS power:
\begin{equation}
   r_k \;=\; r_k^{(\mathrm{p})} \;+\; r_k^{(\mathrm{s})} \;+\; r_k^{(\mathrm{m})}.
   \label{total_reward}
\end{equation}
Here each term targets one of the three goals as discussed further.

\emph{(i) Reduce battery power:} The instantaneous BESS power,
which drives both the converter duty and the cycling degradation, is penalized as
\begin{equation}
    r_k^{(\mathrm{p})} \;=\; -\,w_\mathrm{p}\,
    \bigg(\frac{|P_\mathrm{batt}[k]|}{P_\mathrm{batt}^\mathrm{max}}\bigg)^{\!2},
    \qquad w_\mathrm{p} \geq 0. \label{power_reward}
\end{equation}
The quadratic penalty discourages large actions while keeping the BESS near zero when compensation is unnecessary. 

\emph{(ii) Reduce SoC deviation:} To limit the SoC drift of the BESS, deviation from a reference energy state is penalized by
\begin{equation}
    r_k^{(\mathrm{s})} \;=\; -\,w_\mathrm{s}\,
    \big(\,E_{\mathrm{batt}}[k] - E^{\mathrm{ref}}_{\mathrm{batt}}\,\big)^{2},
    \qquad w_\mathrm{s} \geq 0, \label{soc_reward}
\end{equation}
with $E^{\mathrm{ref}}_{\mathrm{batt}}$ the reference SoC. Maintaining the battery near this reference reduces SoC drift while preserving charging and discharging headroom for subsequent PV ramps.

\emph{(iii) Reduce the maximum power:} A penalty is applied whenever the current action establishes a new running peak:

{\footnotesize
\begin{equation}
    r_k^{(\mathrm{m})} \;=\; -\,w_\mathrm{m}\,
    \frac{\max\!\big\{\,0,\;|P_\mathrm{batt}[k]| - P_\mathrm{batt}^\mathrm{peak}[k]\,\big\}}
         {P_\mathrm{batt}^\mathrm{max}},
    \qquad w_\mathrm{m} \geq 0, \label{max_power_reward}
\end{equation}}with $P_\mathrm{batt}^\mathrm{peak}[k]$ updated by {\footnotesize $P_\mathrm{batt}^\mathrm{peak}[k+1] = \max\{P_\mathrm{batt}^\mathrm{peak}[k],\, |P_\mathrm{batt}[k]|\}$}. The peak-setting events are rare, so a quadratic excess would give a negligible gradient for moderate new peaks, whereas a linear penalty yields a uniformly strong signal whenever a peak is set. Since the running peak fixes the rated power, $w_\mathrm{m}$ governs the trade-off between installed power, and hence capital cost, and steady-state cycling. The reward defined in this section  specifies the per-step preference within the feasibility set. The policy that maximizes its expected discounted return is obtained using proximal policy optimization, as described in the following subsection.

\subsection{Proximal Policy Optimization}
\label{sec: Proximal Policy Optimization}

The CARL controller selects a continuous BESS power action at each time step, and each decision can influence the battery SoC, the available operating flexibility, and the required BESS power and energy capacities over the remainder of horizon. This continuous, long-horizon decision problem motivates the use of an actor--critic framework, where the actor represents the control policy and the critic estimates the expected future return. Accordingly, the proposed policy is trained using proximal policy optimization (PPO)~\cite{schulman2017proximal}, an actor--critic algorithm suitable for continuous action spaces that improves training stability by limiting excessively large policy updates.

\subsubsection{Policy and Value Networks}
For implementation, the state at time step $k$ in~\eqref{state} is represented by a normalized observation vector. In addition to the state variables, deterministic features derived from the feasibility interval and battery energy limits are also included to inform policy about the available operating flexibility. For a PV history length $H$, the observation vector
$o_k \in \mathbb{R}^{H+8}$ is defined as
\begin{equation}
    o_k\;=\; \Big[\;\tfrac{P_\mathrm{pv}[k-H+1]}{P_\mathrm{PV}^\mathrm{rated}},\,\dots,\,\tfrac{P_\mathrm{PV}[k]}{P_\mathrm{PV}^\mathrm{rated}},\;\boldsymbol{\xi}[k]\;\Big]
\end{equation}
where the history is normalized by $P_\mathrm{PV}^\mathrm{rated}$ and the
scalar block is
{\footnotesize
\begin{align}
    \boldsymbol{\xi}[k] = \!\Big(\,\tfrac{P_\mathrm{grid}[k-1]}{P_\mathrm{PV}^\mathrm{rated}},E_{\mathrm{batt}}[k],\,\tfrac{P_\mathrm{batt}^\mathrm{peak}[k]}{P_\mathrm{batt}^\mathrm{max}},\, \,\tfrac{m[k]}{P_\mathrm{batt}^\mathrm{max}},\,\tfrac{w[k]}{P_\mathrm{batt}^\mathrm{max}},\tau_k,\notag \\
    \,\underline{E}_{\mathrm{batt}}-E_{\mathrm{batt}}[k],\,\overline{E}_{\mathrm{batt}}-E_{\mathrm{batt}}[k]\,\Big),.\label{scalar_vector}
\end{align}}
The center and half-width of the feasibility interval are defined as \begin{align}
m[k] = \frac{ \underline{P}_{\mathrm{batt}}[k] + \overline{P}_{\mathrm{batt}}[k] }{2}, \label{feasible_center} \\
w[k] = \frac{ \overline{P}_{\mathrm{batt}}[k] - \underline{P}_{\mathrm{batt}}[k] }{2}, \label{feasible_width}
\end{align} 
respectively. These quantities provide the agent with information on both the location and width of the $\mathcal{F}[k]$. The SoC headroom terms indicate the available margins to the lower and upper limits. The actor $\pi_\theta$ and critic $V_\phi$ are represented by multilayer perceptrons. The actor determines the probability distribution of the normalized action, while the critic estimates the expected future return from current observation. 

\begin{figure*}
    \centering
    \includegraphics[width=0.98\textwidth]{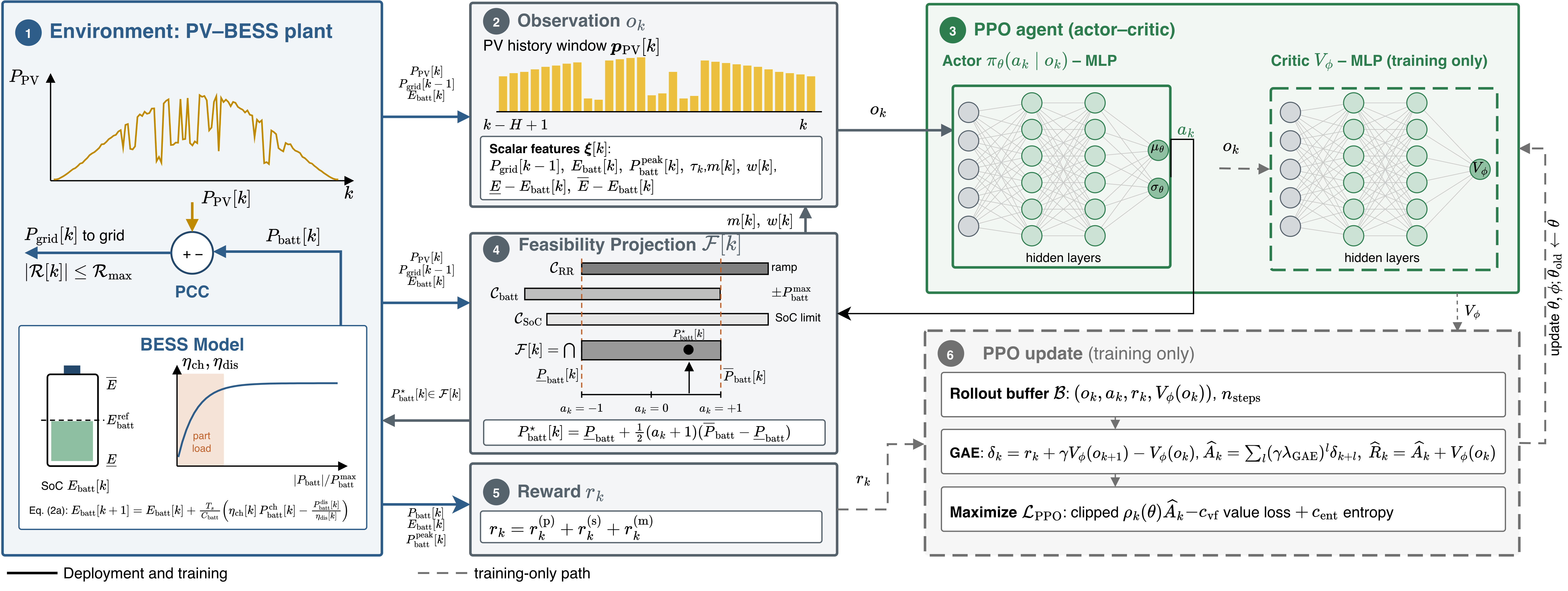}
    \caption{Agent--environment loop of the proposed CARL controller. At each step $k$, the actor $\pi_\theta$ maps the observation $o_k$ to a normalized action $a_k$, which the feasibility projection~\eqref{action_feas} maps onto $P_\mathrm{batt}^\star[k] \in \mathcal{F}[k]$. The critic $V_\phi$ is used only during PPO training. The environment applies $P_\mathrm{batt}^\star[k]$ to the PV-BESS plant dynamics and returns the reward $r_k$~\eqref{total_reward} and the next observation $o_{k+1}$.}
    \label{fig:RLcontroller}
\end{figure*}

\subsubsection{Advantage Estimation and PPO Objective}
The critic estimates the expected cumulative future return from the current observation through the value function
\begin{equation}
   V_{\pi}(o_k) = \mathbb{E}\!\Big[\textstyle\sum_{l=0}^{N-k-1} \gamma^{l}\, r_{k+l}\Big], \qquad
\end{equation}
where $N$ denotes the episode horizon. PPO updates the policy according to the advantage of each sampled action. In this work, the advantage is estimated using generalized advantage estimation (GAE). The temporal difference residual $\delta_k$, measures the discrepancy between the current value estimate and the reward plus the discounted value estimate of the next state, and is defined as,
\begin{equation}
\delta_k=r_k+\gamma V_\phi(o_{k+1})-V_\phi(o_k),
\end{equation}
and the corresponding GAE estimate is
\begin{equation}
\widehat{A}_k=\sum_{l=0}^{N-k-1}(\gamma\lambda_{\mathrm{GAE}})^l\delta_{k+l},
\label{gae}
\end{equation}
where $\lambda_{\mathrm{GAE}}$ controls the bias--variance tradeoff of the advantage estimate. 

Let $\pi_{\theta_{\mathrm{old}}}$ denote the policy used to collect information for an episode and $\pi_\theta$ denote the policy being updated. The probability ratio between the two policies is 
\begin{equation} 
\rho_k(\theta) = \frac{ \pi_\theta(a_k \mid o_k) }{ \pi_{\theta_{\mathrm{old}}}(a_k \mid o_k) }. 
\label{ppo_ratio} 
\end{equation}
PPO constrains the magnitude of each policy update using the clipped surrogate objective
{\footnotesize
\begin{align}
    &\mathcal{L}_\mathrm{PPO}(\theta) \;=\; \mathbb{E}_k\!\Big[\,\min\big(\rho_k(\theta)\widehat{A}_k,\;\mathrm{clip}(\rho_k(\theta),1-\epsilon,1+\epsilon)\widehat{A}_k\big)\,\Big] \notag \\
&\;-\; c_\mathrm{vf}\,\mathbb{E}_k\!\big[(V_\phi(o_k)-\widehat{R}_k)^2\big]
\;+\; c_\mathrm{ent}\,\mathbb{E}_k\!\big[\mathcal{H}(\pi_\theta(\cdot\mid o_k))\big], \label{ppo_obj}
\end{align}}
where $\epsilon$ is the PPO clipping parameter. The clipping operation prevents excessively large changes in the policy during an update. The $c_\mathrm{vf}$ and $c_\mathrm{ent}$ are the value-function and entropy coefficients, respectively. The target return used to train the critic is $\widehat{R}_k = \widehat{A}_k + V_\phi(o_k)$. 
 The first term in~\eqref{ppo_obj} updates the actor while limiting large changes in policy. The second term trains the critic to approximate the expected return, and the entropy term encourages sufficient exploration during training.

 Fig.~\ref{fig:RLcontroller} summarizes the closed-loop operation of the proposed CARL controller. At each time step, the actor selects a normalized action based on the current observation, which is mapped through the feasibility layer to a feasible BESS power setpoint. The resulting setpoint is applied to the PV-BESS system, producing the next system state and reward. The feasibility layer guarantees satisfaction of the operational constraints, while the reward evaluates the selected feasible action according to the BESS sizing objectives. The critic is required only during PPO training, while deployment uses the actor and feasibility mapping.

\subsubsection{Training Procedure}
Each training episode is one calendar day containing $N=1440$ PV measurements at a one-minute sampling resolution. At the beginning of each episode, a day is randomly selected from the training dataset and the BESS SoC is initialized at $E_{\mathrm{batt}}^{\mathrm{ref}}$. Sampling different days exposes the agent to PV trajectories with a wide range of operating and variability conditions. At each time step, the observation vector $o_k$ is constructed. The actor generates a normalized action $a_k$. Using the current state the feasibility interval is computed from~\eqref{feasibility_interval}. The normalized action is subsequently mapped to the feasible BESS power setpoint through~\eqref{action_feas}. The environment applies $P_{\mathrm{batt}}^{\star}[k]$, updates the grid power according to~\eqref{powerflow}, updates the battery energy state using~\eqref{Soc_model}, and evaluates the reward from~\eqref{total_reward}. The resulting transition is stored in the rollout buffer. 

After collecting $n_{\mathrm{steps}}$ transitions, the advantages $\widehat{A}_k$ and target returns $\widehat{R}_k$ are calculated using GAE. The actor and critic are then updated over $K$ minibatch epochs using the PPO objective in~\eqref{ppo_obj}. The updated policy is subsequently used to collect the next rollout. This process continues until the prescribed number of training steps is reached. The discount is set close to one
($\gamma = 0.999$), so that agent remains sensitive to long-horizon consequences such as the SoC depletion caused by pre-positioning errors hours earlier, and $\lambda_\mathrm{GAE} = 0.97$ keeps the advantage estimate low-bias. A constant
entropy coefficient $c_\mathrm{ent} = 10^{-2}$ sustains exploration through the first $10$--$20\%$ of training, before the policy converges on the qualitative timing of charge/discharge events. Algorithm~\ref{alg:ppo} summarizes the training procedure. To evaluate the performance of the proposed controller, the following section presents the benchmark PV smoothing strategies used for comparison.

\begin{algorithm}[t]
\small
\caption{PPO Training}
\label{alg:ppo}
\begin{algorithmic}[1]
\Require Training days $\mathcal{D}_\mathrm{tr}$,
         battery parameters $P_\mathrm{batt}^\mathrm{max}$,
         $C_\mathrm{batt}$, efficiency data $\eta_\mathrm{ch}(\cdot),\eta_\mathrm{dis}(\cdot)$,
         ramp rate limit $\mathcal{R}_\mathrm{max}$,
         reward weights $(w_\mathrm{q},w_\mathrm{c},w_\mathrm{m})$, history length $H$, PPO settings $(\gamma,\lambda_\mathrm{GAE},\epsilon,c_\mathrm{vf},c_\mathrm{ent},n_\mathrm{steps},K)$, total steps $N_\mathrm{tot}$
\Ensure Trained policy $\pi_{\theta}$
\State Initialize actor $\pi_\theta$ and critic $V_\phi$ randomly; $\theta_\mathrm{old}\!\leftarrow\!\theta$
\While{step count $< N_\mathrm{tot}$}
  \State Initialize buffer $\mathcal{B}\leftarrow\varnothing$
  \For{$j=1$ \textbf{to} $n_\mathrm{steps}$}
    \If{episode start}
      \State sample day $d\sim\mathcal{D}_\mathrm{tr}$; set $E_\mathrm{batt}\!\leftarrow\!E_\mathrm{batt}^\mathrm{ref}$,
             $P_\mathrm{grid}[-1]\!\leftarrow\!P_\mathrm{PV}[0]$
    \EndIf
    \State Form observation $o[k]$ from the PV window $\boldsymbol{p}_\mathrm{PV}[k]$ and scalar
           features $\boldsymbol{\xi}[k]$ in \eqref{scalar_vector}
    \State Sample $a_k\sim\pi_{\theta_\mathrm{old}}(\cdot\mid o[k])$
    \State Compute $\mathcal{F}[k]=[\underline{P}_\mathrm{batt}[k],\overline{P}_\mathrm{batt}[k]]$ via \eqref{Feas_interval};
           project $a_k$ to $P_\mathrm{batt}^\star[k]\in\mathcal{F}[k]$ via \eqref{action_feas}
    \State Apply $P_\mathrm{batt}^\star[k]$; update $E_\mathrm{batt}[k{+}1]$ via \eqref{Soc_model} and
           $P_\mathrm{grid}[k]$ via \eqref{powerflow}; compute $r_k$ from \eqref{total_reward}
    \State Store $(o[k],a_k,r_k,V_\phi(o[k]))$ in $\mathcal{B}$
  \EndFor
  \State Compute advantages $\widehat{A}_k$ via GAE$(\gamma,\lambda_\mathrm{GAE})$ and returns
         $\widehat{R}_k=\widehat{A}_k+V_\phi(o[k])$
  \For{$e=1$ \textbf{to} $K$}
    \ForAll{minibatches of $\mathcal{B}$}
      \State Update $\theta,\phi$ by maximizing $\mathcal{L}_\mathrm{PPO}(\theta)$ in \eqref{ppo_obj}
    \EndFor
  \EndFor
  \State $\theta_\mathrm{old}\leftarrow\theta$
\EndWhile
\State \Return $\pi_{\theta}$
\end{algorithmic}
\end{algorithm}

\section{PV Smoothing Strategies}
\label{sec:Smoothing Strategies}
This section presents three benchmark PV smoothing strategies: moving average control (MAC), ramp-rate control (RRC), and an optimization-based approach. These strategies serve as reference methods for evaluating the performance of the proposed CARL controller in terms of BESS power and energy requirements.

\subsection{Moving Average Control (MAC)}    
\label{Sec: Moving Average}
MAC is a filter-based method in which PV power is averaged over a window to generate the grid power reference:
\begin{equation}
P_{\mathrm{grid}}[k]
=
\frac{1}{W}
\sum_{j=k-W+1}^{k} P_{\mathrm{PV}}[j],
\label{ma_grid}
\end{equation}
where the window length $W$ determines the degree of smoothing applied to the PV output. $P_\mathrm{batt}^{\star}[k]$ is obtained from the difference between instantaneous PV power and the smoothed grid reference, 
\begin{equation}
P_\mathrm{batt}^{\star}[k]
=
P_{\mathrm{PV}}[k]-P_{\mathrm{grid}}[k].
\label{ma_batt}
\end{equation}

Conventional MAC formulation does not enforce the RRL explicitly, and compliance is instead achieved indirectly by selecting a window length $W$ such that the filtered grid output satisfies the RRL. A larger window provides stronger attenuation of short-term fluctuations but also increases the temporal lag between the instantaneous PV power and the filtered reference, which must be compensated by the BESS even when the original PV output already satisfies the RRL. This continuous compensation can increase both battery energy throughput and the required BESS capacity~\cite{marcos2014control}. Adaptive MAC methods address some of these limitations by adjusting $W$ according to the RRL and the available BESS operating range~\cite{kini2019mitigation}. However, these methods may still lead to SoC drift due to unnecessary BESS operation. To avoid unnecessary BESS operation associated with MAC, RRC instead activates the battery only when needed to satisfy the RRL.

\subsection{Ramp-Rate Control (RRC)}
\label{sec:APP_RRC}
RRC is a reactive smoothing method that selects the BESS power setpoint closest to zero within the feasibility interval, thereby minimizing unnecessary battery operation while satisfying the RRL and BESS operational constraints. Using the feasibility interval $\mathcal{F}[k]$ derived in Section~\ref{sec: Ramp-Rate constraint section}, the BESS remains idle whenever $0 \in \mathcal{F}[k]$, otherwise, the controller selects the nearest boundary of $\mathcal{F}[k]$,
\begin{equation}
P^{\star}_{\mathrm{batt}}[k] = \max\left\{\underline{P}_{\mathrm{batt}}[k],\min\left\{0,\,\overline{P}_{\mathrm{batt}}[k]\right\}\right\}.
\label{rrc_batt}
\end{equation}
RRC avoids the lag associated with filter-based methods and uses the minimum magnitude BESS action, therefore reducing unnecessary battery cycling. However, its decisions depend only on the current operating condition  and do not account for the effect of current actions on future SoC or ramp events. As a result, the BESS must compensate for full RRL violation as it occurs, which can increase the required BESS power rating. Moreover, Without an SoC management method, repeated charging or discharging during consecutive ramp events can drive the SoC towards its operating limits, reduce the available energy headroom for the future events, and consequently increase the required BESS energy capacity~\cite{huo2020novel}. SoC feedback strategies can partially mitigate these issues, but RRC methods inherently trade off ramp-rate compliance, battery capacity requirements, and cycling-induced degradation~\cite{gonzalez2023influence,mao2025overview}.

\subsection{Input Convex Neural Network (ICNN) Optimization}
\label{sec:ICNN_opt}
Unlike MAC and RRC methods, the optimization-based methods can explicitly incorporate BESS sizing objectives in the PV smoothing problem while enforcing the RRL and BESS operational constraints. It formulates the problem over the full daily horizon with perfect foresight of the PV power profile, thereby providing a noncausal optimal reference against which the performance of the other controllers can be evaluated. Direct incorporation of the nonlinear part-load efficiency model, described in Section~\ref{sec:BESS model}, would lead to a nonconvex optimization problem, leading to local optimal solutions. Conversely, assuming constant efficiencies would introduce errors in the battery SoC trajectory and could affect the estimated BESS requirements. To model nonlinear efficiencies while maintaining a convex optimization formulation, this study adopts the input-convex neural network (ICNNs) BESS model developed in~\cite{omidi2025towards}, in which the charging and discharging part in~\eqref{Soc_model} are approximated by ICNNs.

The optimization minimizes the peak BESS power and penalizes the deviation of the SoC from its reference value over the daily horizon, with the objective function given by

{\small
\begin{equation}
\min \quad
\left\|P_\mathrm{batt}^\mathrm{ch}
-P_\mathrm{batt}^\mathrm{dis}\right\|_{\infty}
+\lambda_{\mathrm{soc}}
\left\|\boldsymbol{E}_{\mathrm{batt}}
-E_{\mathrm{batt}}^{\mathrm{ref}}\right\|_{2}^{2},
\label{opt_main_obj}
\end{equation}}here $\lambda_{\mathrm{soc}}$ weights the SoC deviation penalty. The optimization is subject to the grid power balance, RRL, BESS power and SoC constraints, nonlinear efficiency dynamics represented by the ICNNs, and a terminal SoC condition. The ICNN activation functions and the constraint preventing simultaneous charging and discharging are represented using mixed-integer formulations. The complete mathematical formulation is provided in Appendix~\ref{app:icnn_optimization}.

Since the optimization uses the complete daily PV trajectory, it can adjust the BESS SoC in anticipation of future ramp events while satisfying the operational constraints and restoring the SoC to its reference value at the end of the horizon. Therefore, the resulting solution serves as a perfect-foresight benchmark for evaluating the BESS power and energy requirements of the causal smoothing controllers.

\section{Simulation Results and Case Studies}
\label{sec:simulation and results}
This section evaluates the proposed CARL controller and estimates the BESS size required  to maintain compliance with the grid-code RRL. The controller is trained and tested using historical PV data from three geographically distinct sites with different levels of PV power variability and plant capacity. Its performance is benchmarked against the MAC, RRC and ICNN methods presented in Section~\ref{sec:Smoothing Strategies}. For all controllers, the BESS SoC is initialized at its reference value at the beginning of each daily simulation, assuming that any SoC deviation from the previous day is restored overnight to ensure sufficient energy availability for the next day. The system parameters and PPO hyperparameters used in the numerical simulation are summarized in Table~\ref{tab:parameters}. For each dataset, the reward weights were chosen using a two-stage grid search. First, a broad range of values was tested to identify suitable ranges for each weight. Then, three candidate values were selected for each of the three reward terms, and all 27 possible combinations were evaluated. Each configuration was trained using PPO and compared based on maximum BESS power and SoC deviation of the training data. The selected configuration was retrained using three independent random seeds, and the resulting performance metrics were averaged to assess the consistency of the results. 

\begin{table}[t]
\centering
\caption{System and training parameters used in the numerical simulations.}
\label{tab:parameters}
\footnotesize

\begin{threeparttable}
\begin{tabular*}{0.95\columnwidth}
{@{\extracolsep{\fill}} l c c @{}}
\toprule
\textbf{Parameter} & \textbf{Symbol} & \textbf{Value} \\
\midrule

\multicolumn{3}{l}{\textit{System and BESS parameters}} \\
Sampling interval & $\Delta t$ & 1~min \\
Episode horizon & $N$ & 1440 steps \\
Minimum SoC & $\underline{E}_{\mathrm{batt}}$ & 0.05 \\
Maximum SoC & $\overline{E}_{\mathrm{batt}}$ & 0.95 \\
Reference/initial SoC & $E_{\mathrm{batt}}^{\mathrm{ref}}$ & 0.5 \\
Maximum charging efficiency & $\eta_{\mathrm{ch}}^{\mathrm{max}}$ & 0.93 \\
Minimum discharging efficiency & $\eta_{\mathrm{dis}}^{\mathrm{min}}$ & 0.5 \\

\midrule
\multicolumn{3}{l}{\textit{PPO hyperparameters}} \\
PV history length
    & $H$ & 30 \\
Discount factor
    & $\gamma$ & 0.999 \\
GAE parameter
    & $\lambda_{\mathrm{GAE}}$ & 0.97 \\
PPO clipping parameter
    & $\epsilon$ & 0.2 \\
Value-function coefficient
    & $c_{\mathrm{vf}}$ & 0.5 \\
Entropy coefficient
    & $c_{\mathrm{ent}}$ & 0.01 \\
Rollout length
    & $n_{\mathrm{steps}}$ & 4096 \\
Minibatch size
    & $n_{\mathrm{batch}}$ & 256 \\
Learning rate
    & $\alpha$ & $3\times10^{-4}$ \\
Total training steps
    & $N_{\mathrm{tot}}$ & 3M\\
\bottomrule
\end{tabular*}

\end{threeparttable}
\end{table}

\subsection{PV Datasets and Site Characteristics}
\label{sec:dataset}
The proposed controller is evaluated on three independent PV datasets collected from geographically distinct sites: Site~A located in Burlington~\cite{Omidi2026HES}, Vermont, Site~B located in Shungnak, Alaska, and Site~C located in Rutland, Vermont. Sites~A and B employ bifacial PV modules, whereas Site~C uses monofacial modules. In addition, Site~A is equipped with module-level power optimizers. These technologies influence the aggregate plant response to irradiance fluctuations and can therefore affect the observed ramp-rate variability. The sites differ in plant capacity, providing a diverse set of operating conditions for evaluating the proposed controller. The Site~A dataset contains eight months of measured AC-side PV power data, whereas Sites~B and C datasets each contain one calendar year of data. All datasets are sampled at $\Delta t = 1$~min. For each site, the days in every month were randomly divided into training and test sets using a 70/30 \% split, respectively. Since a training episode is a single calendar day, this day-level random split exposes the policy to the full seasonal range of irradiance conditions while keeping the test days fully disjoint from training.

\begin{figure}
  \centering
  \begin{subfigure}{\linewidth}
    \includegraphics[width=0.9\linewidth]{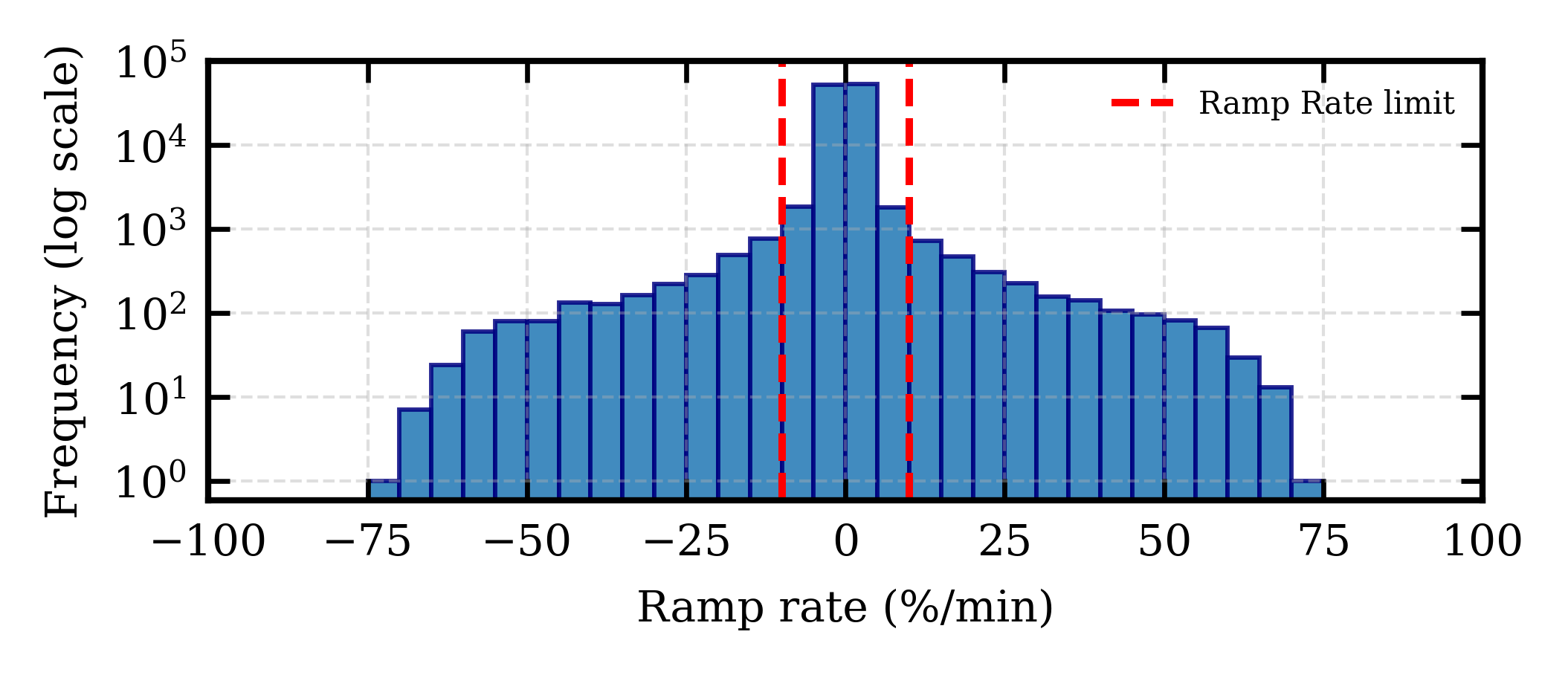}
    \caption{Site A}\label{fig:rr_a}
  \end{subfigure}\\
  \begin{subfigure}{\linewidth}
    \includegraphics[width=0.9\linewidth]{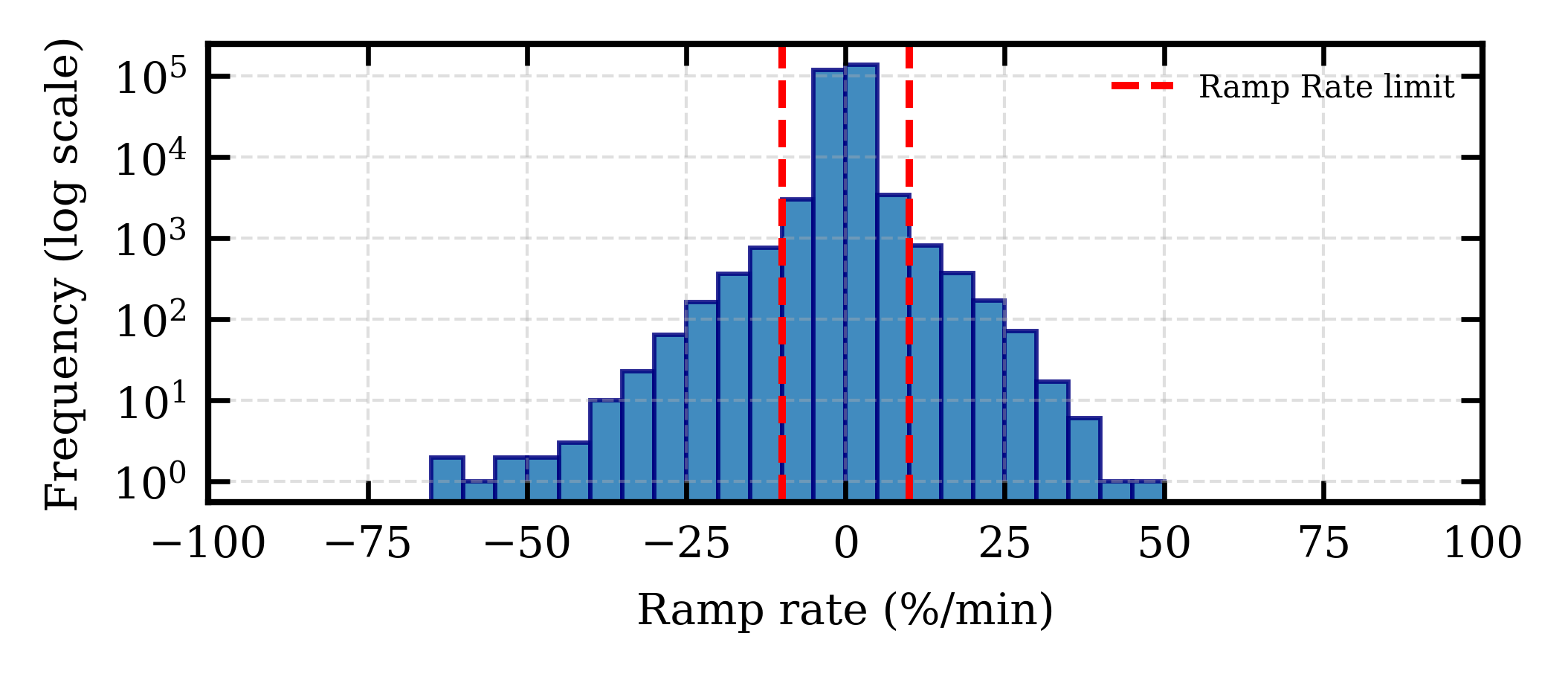}
    \caption{Site B}\label{fig:rr_b}
  \end{subfigure}\\
  \begin{subfigure}{\linewidth}
    \includegraphics[width=0.9\linewidth]{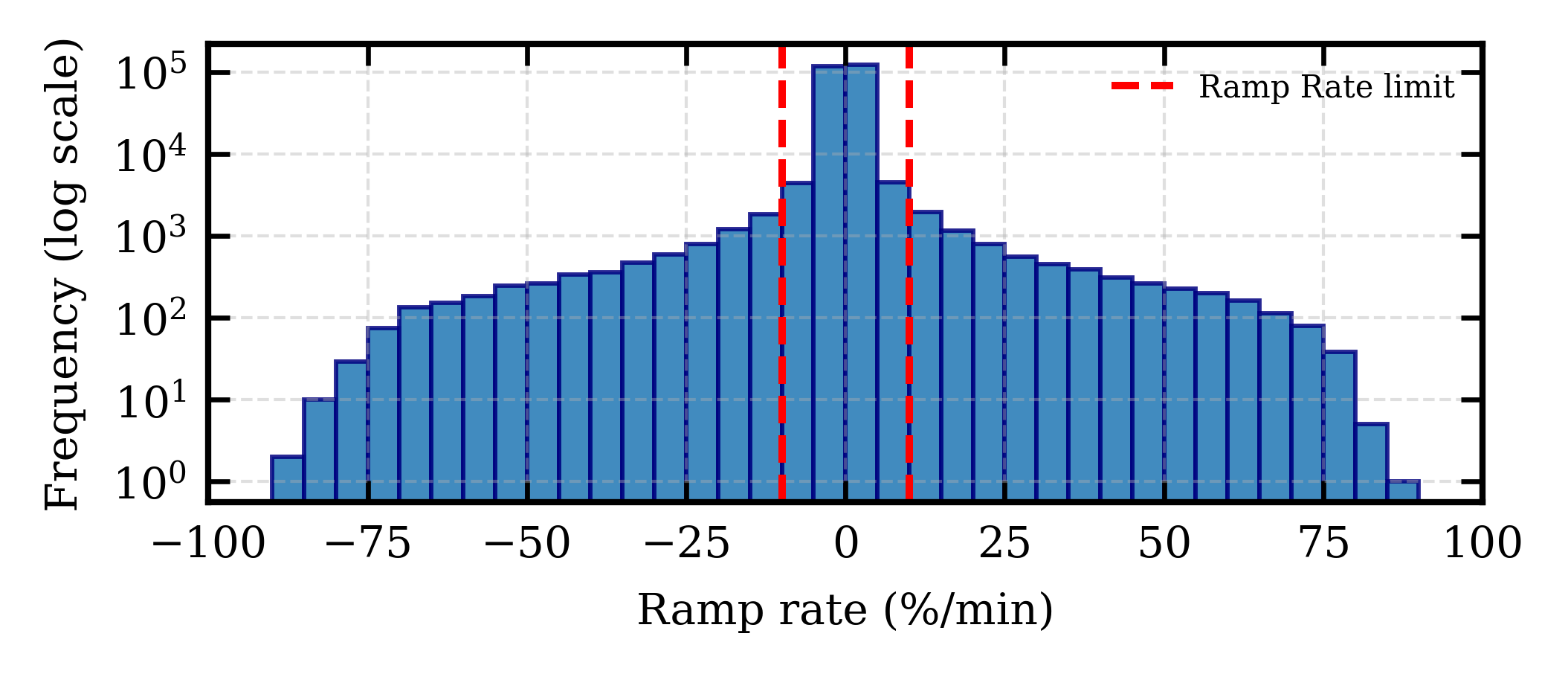}
    \caption{Site C}\label{fig:rr_c}
  \end{subfigure}
\caption{Ramp-rate distributions of the raw PV output for the three datasets: (a) Site~A, (b) Site~B, and (c) Site~C. The distributions are computed from~\eqref{ramp_rate_equation} over daylight samples and shown as probability densities on a logarithmic scale. The vertical dashed lines indicate the grid-code limits of $\pm 10\,\%/\mathrm{min}$.}
  \label{fig:rr_dist}
\end{figure}

For each site, the ramp rate of the raw PV power was calculated at each daylight time step, and expressed in \%/min. The resulting ramp-rate distributions are shown in Fig.~\ref{fig:rr_dist}. Table~\ref{tab:ramp_stats} summarizes the key characteristics and ramp-rate statistics of the three PV datasets. The datasets exhibit different levels of output variability. Site~C shows the highest variability, with a ramp-rate standard deviation of $\sigma({\lvert\mathcal{R}[k]\rvert})=6.9\%/\mathrm{min}$, a $99^{\mathrm{th}}$-percentile ramp-rate magnitude of $40.4\%/\mathrm{min}$, and the largest proportion of daylight samples exceeding the grid-code RRL, at 5.1\%.

\begin{table}[t]
  \centering
  \caption{Ramp-rate statistics of the three distinct datasets (raw PV output,
           daylight samples, $\Delta t = 1$~min).}
  \label{tab:ramp_stats}
  \renewcommand{\arraystretch}{1.15}
  \begin{tabular}{l ccc}
    \toprule
     & \multicolumn{3}{c}{Dataset} \\
    \cmidrule(lr){2-4}
    Statistic & Site A & Site B & Site C \\
    \midrule
    Plant Capacity (kW)                            & 6.1   & 200    & 510   \\
    Mean $ |\mathcal{R}[k]|$    (\%/min)                & 1.8   & 0.8    & 2.1   \\
    Mean $ |\mathcal{R}[k]|>10$  (\%/min)             & 24.6   & 16    & 27.2   \\
    $\sigma_{ |\mathcal{R}[k]|}$ (\%/min)                 & 5.7   & 2.1    & 6.9   \\
    Max up-ramp (\%/min)                        & 71.7   & 45.8    & 87   \\
    Max down-ramp (\%/min)                        & 71.6   & 61.7    & 88   \\
    $99^{\text{th}}$$ |\mathcal{R}[k]|$   (\%/min)            & 33.1   & 10.5   & 40.4   \\
    Steps with $|\mathcal{R}[k]|>10$ (\%) & 4.2   &1.1    & 5.1   \\
    \bottomrule
  \end{tabular}
\end{table}

\subsection{Comparative Performance on a Representative Day}
\label{sec:representative_days}
\begin{figure}
  \centering
  \begin{subfigure}{\linewidth}
    \includegraphics[width=0.9\linewidth]{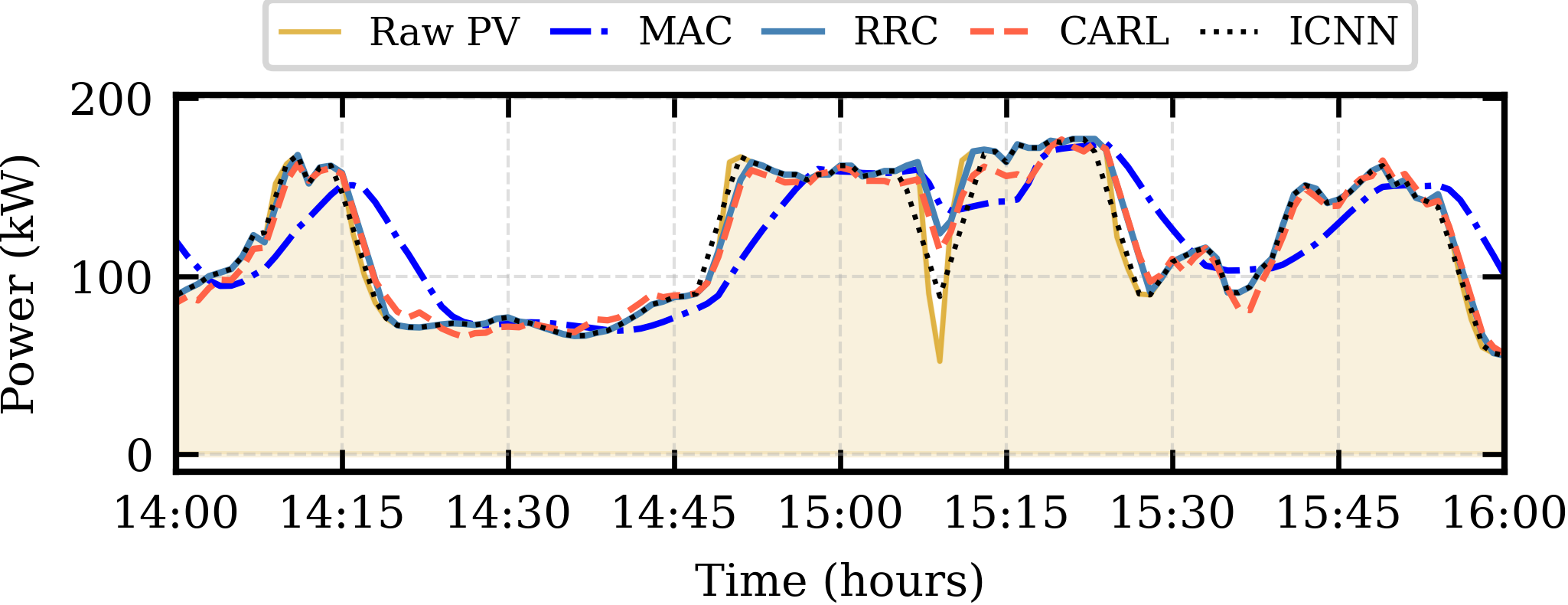}
    \caption{}\label{fig:rep_day_a}
  \end{subfigure}\\
  \begin{subfigure}{\linewidth}
    \includegraphics[width=0.9\linewidth]{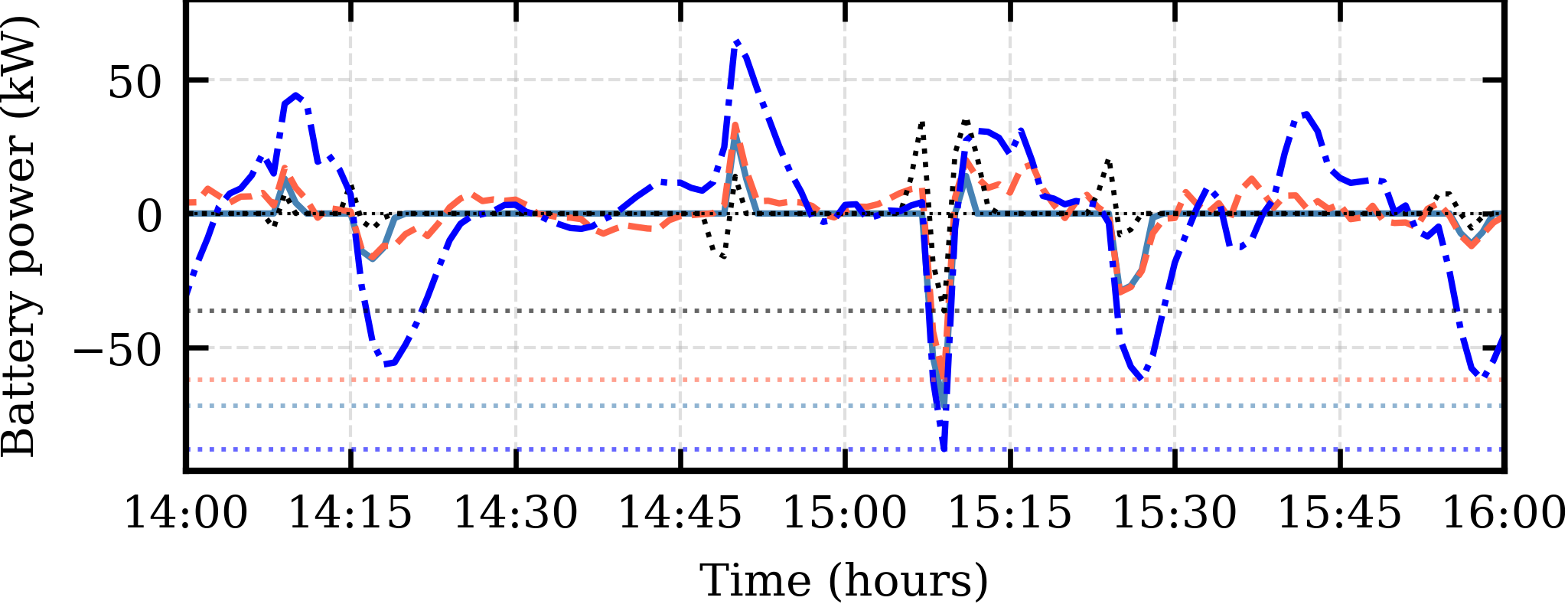}
    \caption{}\label{fig:rep_day_b}
  \end{subfigure}\\
  \begin{subfigure}{\linewidth}
    \includegraphics[width=0.9\linewidth]{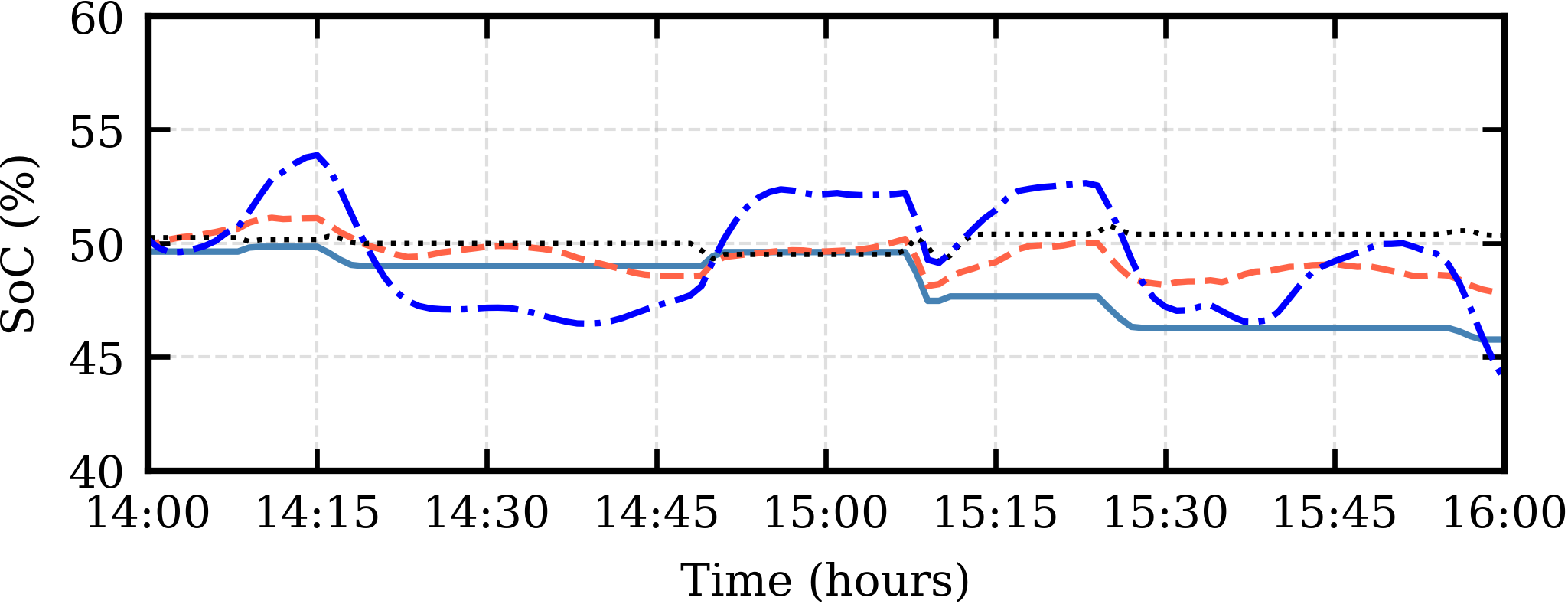}
    \caption{}\label{fig:rep_day_c}
  \end{subfigure}
  \begin{subfigure}{\linewidth}
    \includegraphics[width=0.9\linewidth]{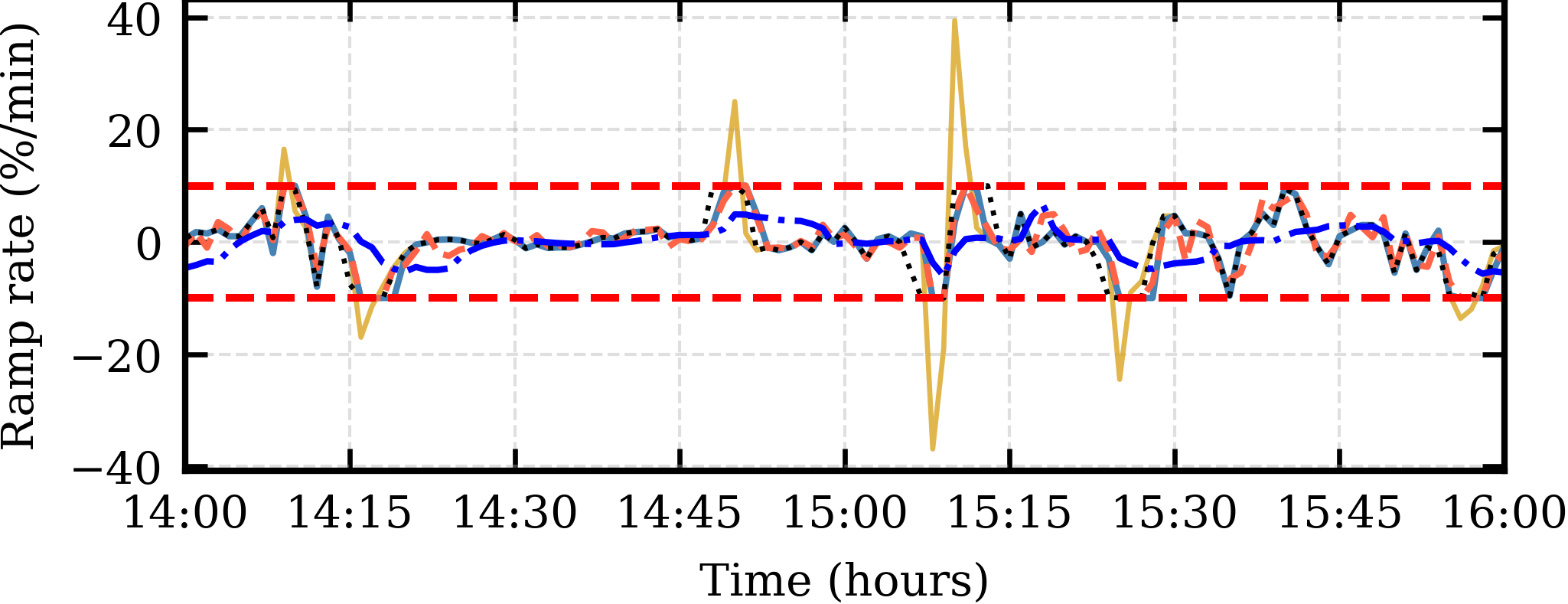}
    \caption{}\label{fig:rep_day_d}
  \end{subfigure}
  \caption{Comparative operation of the MAC, RRC, CARL, and ICNN controllers on a representative test day at Site~B: (a) smoothed power output $P_\mathrm{grid}$ at the PCC versus the raw PV output $P_\mathrm{PV}$; (b) dispatched battery power $P_\mathrm{batt}$; (c) state of charge $\mathrm{SoC}$; (d) per-step ramp rate $\mathcal{R}$ of each smoothed output against the grid-code band $\mathcal{R}_\mathrm{max}=\pm10\,\%/\min$ (dashed lines).}
  \label{fig:rep_day_compare}
\end{figure}

The performance of the proposed CARL controller is compared with the RRC, MAC, and ICNN optimization controllers on the Site~B test day associated with the maximum BESS power demand. Fig.~\ref{fig:rep_day_compare} presents a two-hour snapshot from this day and compares the controllers in terms of the smoothed grid output, $P_\mathrm{grid}$, at the PCC against the raw PV output, $P_\mathrm{PV}$ (Fig.~\ref{fig:rep_day_a}), the dispatched battery power, $P_\mathrm{batt}$ (Fig.~\ref{fig:rep_day_b}), the resulting SoC (Fig.~\ref{fig:rep_day_c}), and the per-step ramp rate, $\mathcal{R}$, of each smoothed output (Fig.~\ref{fig:rep_day_d}).
 
All controllers maintain $\mathcal{R}[k]$ within the $\pm10\,\%/\min$ limit throughout the test day. The MAC forces the $P_\mathrm{grid}$ to follow a moving average of the raw PV output. Therefore, the BESS must compensate for the difference between the raw PV power and its lagged average. During a rapid PV power change, this difference can approach the full magnitude of the PV power change, causing MAC to produce the largest peak BESS power. The RRC responds only when a ramp-rate violation occurs and applies the minimum corrective action required to restore compliance. The proposed CARL controller achieves a lower peak BESS power than the RRC and MAC by initiating charging or discharging before the ramp events. This behavior is consistent with the long-horizon discount factor ($\gamma=0.999$), and the running-peak penalty in~\eqref{max_power_reward}. In addition, the SoC-centering penalty in~\eqref{soc_reward} maintains the SoC near its reference midpoint, as shown in Fig.~\ref{fig:rep_day_c}. The ICNN controller provides the optimal solution to the formulated smoothing problem under perfect knowledge of the PV profile. Using this future information, it adjusts the BESS power and SoC in anticipation of upcoming ramp events, thereby reducing the peak BESS power requirement and maintaining the SoC within a narrower operating range than the other controllers. These results indicate that the proposed CARL controller provides effective ramp-rate mitigation while reducing the peak BESS power and energy relative to the RRC and MAC controllers on the representative test day. Aggregated performance over the test dataset of site~B is quantified in the following subsection.

\subsection{BESS Sizing}

This subsection evaluates the BESS power and energy requirements associated with each controller over the entire site~B test dataset. For each controller and ramp-rate limit, the required peak BESS power, $P_\mathrm{batt}$, and maximum energy capacity, $C_\mathrm{batt}$, are determined by applying the sizing procedure in Algorithm~\ref{alg:sizing}. To enable a consistent comparison, the BESS power and energy capacity requirements are normalized by the rated PV plant capacity.
\label{sec:power_energy_crate}

\begin{figure}
  \centering
  \begin{subfigure}{\linewidth}
    \includegraphics[width=\linewidth]{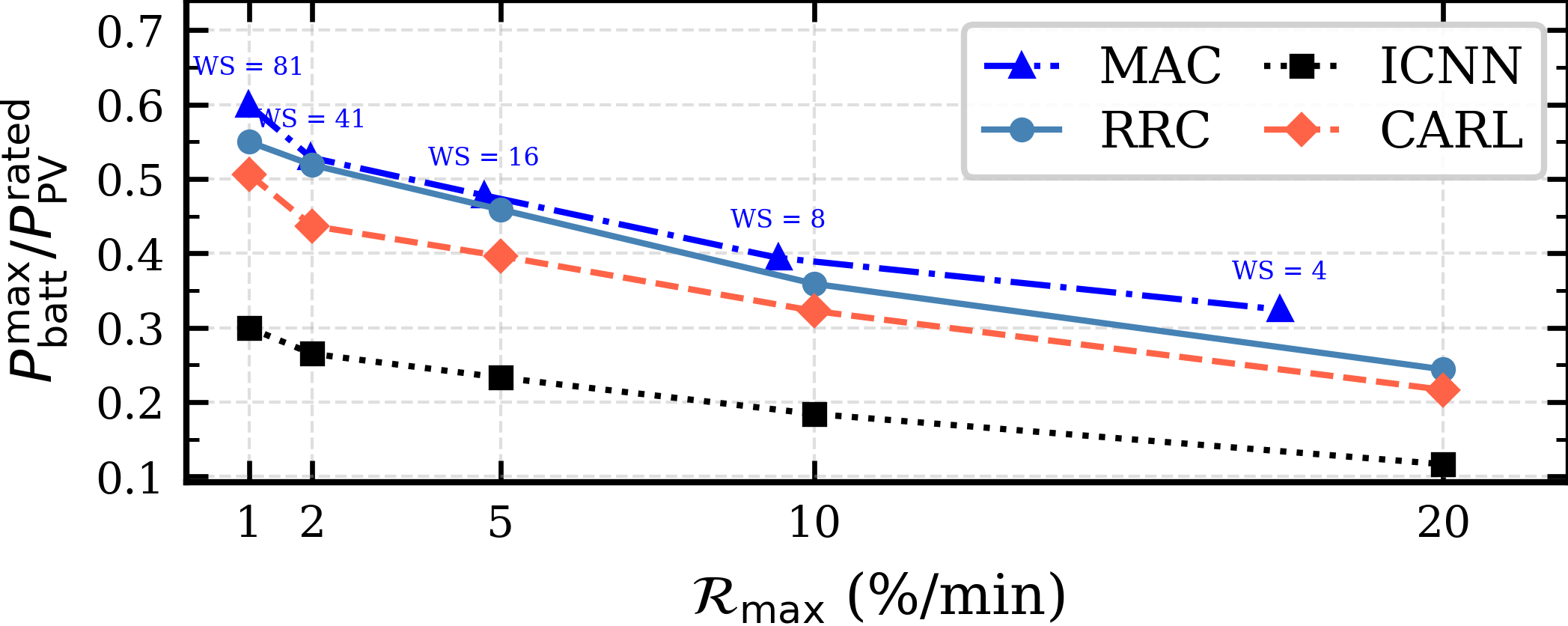}
    \caption{}\label{fig:power_size}
  \end{subfigure}\\
  \begin{subfigure}{\linewidth}
    \includegraphics[width=\linewidth]{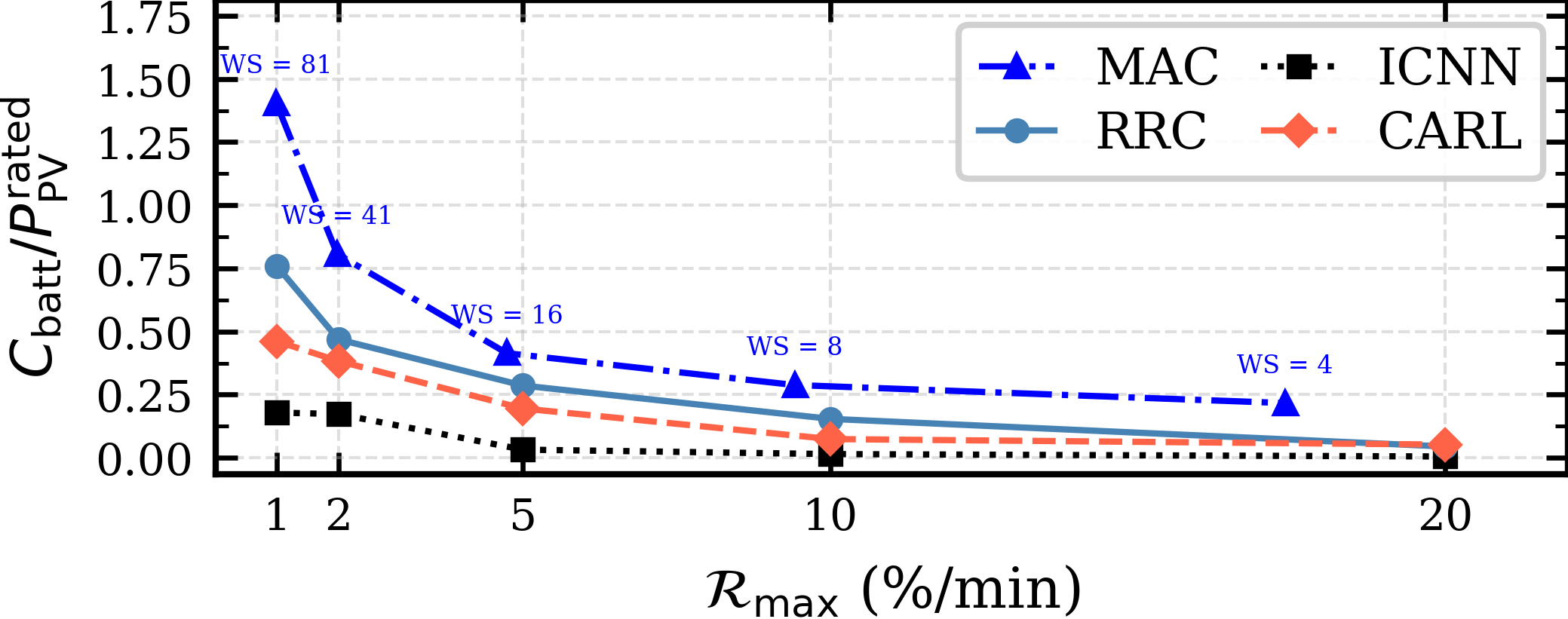}
    \caption{}\label{fig:energy_size}
  \end{subfigure}\\
  \caption{Battery sizing versus ramp-rate limit $\mathcal{R}_\mathrm{max}$ for MAC, RRC, proposed CARL, and ICNN optimization controllers over the entire test dataset of Site~B: (a) normalized required peak BEES power $P_\mathrm{batt}^\mathrm{max}$; (b) normalized required BEES energy capacity $C_\mathrm{batt}$. MAC is swept over its averaging window rather than $\mathcal{R}_\mathrm{max}$, so each MAC point is plotted at its achieved ramp rate and labeled with the window size (WS).}
  \label{fig:power_energy_size}
\end{figure}

\subsubsection{Peak BESS Power}
Fig.~\ref{fig:power_size} shows the required peak BESS power, $P_\mathrm{batt}^\mathrm{max}$, as a function of $\mathcal{R}_\mathrm{max}$. For all four controllers, the required peak power decreases as the ramp-rate constraint is relaxed because a large fraction of PV power fluctuations can be passed directly to the grid. The ICNN optimization controller requires the lowest peak power across the entire range by exploiting full-horizon PV information and directly minimizing peak battery power. It serves as an effective lower bound for the peak BESS power. The relative performance of MAC and RRC depends on both the RRL and PV power profile and variability. For the site~B PV power profile, MAC results in a higher peak BESS power requirement than RRC.  As $\mathcal{R}_{\max}$ is relaxed, the required RRC power decreases rapidly because less intervention is needed to satisfy the ramp-rate constraint, whereas MAC continues to compensate for the difference between the instantaneous PV output and its trailing average. The proposed CARL controller requires consistently lower peak BESS power than both RRC and MAC over the considered range. Relative to RRC, CARL reduces the required peak power by approximately $7.9$ to $15.8\%$ over the entire range. Compared with MAC, the reduction is more pronounced, ranging from approximately $15.6\%$ to $33.3\%$. At $\mathcal{R}_{\max}=10\%/\min$, for example, the required peak power decreases from $0.36$~p.u. for RRC to $0.32$~p.u. for CARL. The observed reduction is consistent with the running-peak penalty in~\eqref{max_power_reward} and the long discount horizon, which encourage the controller to distribute battery action over multiple time steps rather than concentrate it at the occurrence of a ramp event.

\subsubsection{BESS Energy Span}
Fig.~\ref{fig:energy_size} presents the required BESS energy capacity, $C_\mathrm{batt}$, normalized by the rated PV power as a function of $\mathcal{R}_\mathrm{max}$. The ICNN controller requires the smallest energy capacity over the entire range. By exploiting perfect foresight, it pre-positions the battery SoC before anticipated ramp events and maintain the SoC near its midpoint, thereby reducing the required SoC deviation and energy capacity. Among the causal controllers, MAC requires the largest energy capacity, particularly under tighter ramp-rate constraints, because it continuously compensates for the difference between the instantaneous PV power and its trailing average, even when no ramp-rate violation occurs.

The proposed CARL controller requires the lowest energy capacity among the causal methods for $\mathcal{R}_{\max}=1$, $2$, $5$, and $10,\%/\mathrm{min}$, which is consistent with the SoC-centering penalty in~\eqref{soc_reward}. At $\mathcal{R}_{\max}=10\%/\mathrm{min}$, CARL reduces the required energy capacity by approximately $51.9\%$ relative to RRC and $74.3\%$ relative to MAC. As the ramp-rate limit is further relaxed, however, the energy advantage of CARL relative to RRC diminishes and eventually reverses. At $\mathcal{R}_{\max}=20,\%/\mathrm{min}$, CARL requires $0.052$~p.u.-h, whereas RRC requires only $0.045$~p.u.-h. This behavior can be attributed to CARL continuing to position and utilize the battery in anticipation of possible ramp events, while RRC acts only when required to enforce the RRL and otherwise allows the battery to remain nearly idle. Therefore, the energy-sizing benefit of the proposed CARL controller relative to conventional RRC is most pronounced under tighter ramp-rate constraints. At $\mathcal{R}_{\max}=20\,\%/\mathrm{min}$, the CARL controller requires $0.2165$~p.u. of peak power but only $0.052$~p.u.-h of energy capacity, corresponding to a C-rate of approximately $4.16$C. This high C-rate indicates that, under relaxed ramp-rate constraints, the BESS sizing is governed more strongly by the required power rating than by the energy capacity.

\subsubsection{Controller Dispatch Characteristics}
\label{sec:dispatch_characteristics}
\begin{figure}
    \centering
    \includegraphics[width=\columnwidth]{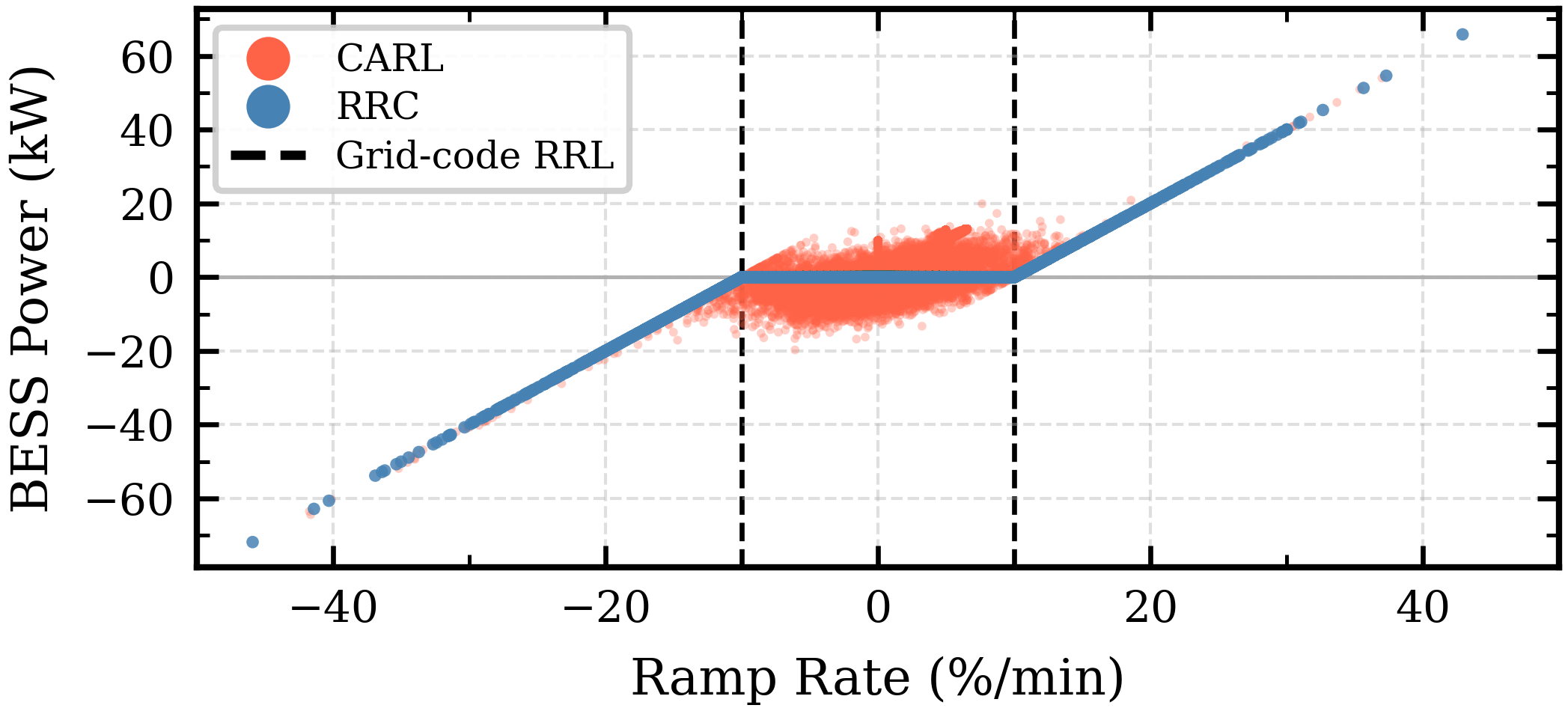}
    \caption{Dispatched BESS power versus the per-step ramp of the PV output over all daylight samples of the Site~B test set, for the RRC and CARL controllers. Vertical dashed lines mark the grid-code RRL $\mathcal{R}_{\max}=\pm10\,\%/\min$.}
    \label{fig:batt_vs_ramp}
\end{figure}

To better understand the operating behavior of the proposed CARL controller, its battery dispatch is compared with that of RRC controller as a function of per-step PV ramp rate. Fig.~\ref{fig:batt_vs_ramp} shows this relationship for all daylight samples of the Site~B test dataset. RRC exhibits a clear dead band, with zero battery action inside the RRL and approximately linear corrective action outside it, reflecting its reactive operation. In contrast, CARL dispatches non-zero battery power even within the $\mathcal{R}_\mathrm{max}$ limit and produces a range of actions for the same effective ramp because its decision depends on the full policy state in~\eqref{state}, including SoC, recent PV history, running peak power, and time. This allows CARL to distribute battery action over multiple steps and reduce extreme power and energy demands while maintaining ramp-rate compliance through the feasibility projection in~\eqref{action_feas}.

\subsection{BESS Sizing and Cost Estimation Across Multiple PV Sites}

\begin{table*}
  \centering
  \caption{Required battery power $P_\mathrm{rated}$ (kW), energy
  $C_\mathrm{rated}$ (kWh), implied C-rate, and estimated capital cost (\$) using
  the 2024 base-year BESS cost reported in~\cite{cole2025cost},
  for each controller at Sites~A--C and
  $\mathcal{R}_\mathrm{max}=10\,\%/\min$.}
  \label{tab:power_energy_cost_vs_duration}
  
  \renewcommand{\arraystretch}{1.3}
  \resizebox{\textwidth}{!}{%
  \begin{tabular}{l|c c c c|c c c c|c c c c}
    & \multicolumn{4}{c|}{\textbf{Site~A}}
    & \multicolumn{4}{c|}{\textbf{Site~B}}
    & \multicolumn{4}{c}{\textbf{Site~C}} \\
    \cline{2-13}
    
    \textbf{Model}
    & $P_\mathrm{rated}$ & $C_\mathrm{rated}$ & C-rate & Cost
    & $P_\mathrm{rated}$ & $C_\mathrm{rated}$ & C-rate & Cost
    & $P_\mathrm{rated}$ & $C_\mathrm{rated}$ & C-rate & Cost \\
    
    & (kW) & (kWh) & ($\mathrm{h}^{-1}$) & (\$)
    & (kW) & (kWh) & ($\mathrm{h}^{-1}$) & (\$)
    & (kW) & (kWh) & ($\mathrm{h}^{-1}$) & (\$) \\
    \hline
    
    \textbf{MAC}
    & 3.7 & 1.6 & 2.31 & 1{,}762
    & 78.8 & 57.6 & 1.37 & 43{,}195
    & 384 & 194 & 1.98 & 189{,}602 \\
    
    \textbf{RRC}
    & 3.5 & 1.2 & 2.92 & 1{,}591
    & 72 & 31 & 2.32 & 34{,}255
    & 378 & 198 & 1.91 & 188{,}334 \\
    
    \textbf{CARL (proposed)}
    & \textbf{3.14} & \textbf{0.63} & 4.98 & \textbf{1{,}320}
    & \textbf{64.5} & \textbf{15} & 4.30 & \textbf{27{,}609}
    & \textbf{345} & \textbf{74} & 4.66 & \textbf{146{,}174} \\
\hline 
    \textbf{ICNN Optimization}
    & 1.7 & 0.3 & 5.67 & 705
    & 37 & 5 & 7.40 & 14{,}969
    & 189 & 46 & 4.11 & 81{,}394 \\
    
  \end{tabular}}
\end{table*}

To evaluate the performance of the proposed CARL framework across different PV operating conditions, all four controllers were independently evaluated on the three datasets described in Section~\ref{sec:dataset}. Table~\ref{tab:power_energy_cost_vs_duration} summarizes the resulting BESS power and energy ratings, implied C-rate, and estimated capital costs for a RRL of $\mathcal{R}_\mathrm{max}=10\,\%/\min$. The BESS capital cost is estimated using
\begin{equation}
    \text{Cost} = P_\mathrm{rated}\times\$372/\mathrm{kW} + C_\mathrm{rated}\times\$241/\mathrm{kWh},
    \label{eq:cost_split}
\end{equation}
with {\small $P_\mathrm{rated}=\max(P_\mathrm{batt}^\mathrm{max},\,C_\mathrm{rate}\,C_\mathrm{batt})$} and {\small $C_\mathrm{rated}=P_\mathrm{rated}/C_\mathrm{rate}$}~\cite{cole2025cost}. Across all three sites, CARL requires lower peak BESS power and energy capacity than MAC and RRC, while the ICNN optimization solution provides the lowest requirements. Compared with RRC, CARL reduces the peak power requirement by approximately $10.3\%$, $10.4\%$, and $8.7\%$ at Sites~A, B, and C, respectively. The corresponding reductions in required energy capacity are approximately $47.5\%$, $51.6\%$, and $62.6\%$, respectively.

For the considered ramp-rate limit of $\mathcal{R}_{\max}=10\%/\min$, the implied C-rate is greater than unity for all datasets and control methods, indicating that PV smoothing is predominantly power-constrained rather than energy-constrained under this operating condition. In other words, the required peak BESS power is relatively large compared with the required energy capacity, although the exact power-to-energy ratio depends on the PV variability and the control strategy. This result has an important implication for practical BESS sizing. Utility-scale BESS installations are commonly designed with fixed and relatively low C-rates~\cite{eia860_2024}. When the available BESS C-rate is lower than the C-rate implied by the PV-smoothing requirement, the peak power requirement becomes the decisive sizing constraint. Therefore, for a fixed C-rate BESS, reductions in peak power directly reduce both the required BESS rating and the associated capital cost. This behavior is consistent with the running-peak penalty in~\eqref{max_power_reward}, which explicitly discourages large BESS power.

\subsection{Cross-Dataset Evaluation}
\begin{table}[t]
\centering
\caption{Cross-dataset generalization performance of CARL. Rows indicate the training dataset and columns indicate the testing dataset.}
\label{tab:cross_dataset_generalization}
\renewcommand{\arraystretch}{1.25}

\begin{subtable}[t]{0.95\columnwidth}
\centering
\caption{Normalized peak BESS power,
$P_{\mathrm{batt}}^{\max}/P_{\mathrm{PV}}^{\mathrm{rated}}$.}
\label{tab:cross_dataset_power}
\begin{tabular}{c|ccc}
\hline
\textbf{Train $\backslash$ Test}
& \textbf{Site A}
& \textbf{Site B}
& \textbf{Site C} \\
\hline
\textbf{Site A} & \cellcolor{gray!25}0.51 & 0.35 & 0.71 \\
\textbf{Site B} & 0.56 & \cellcolor{gray!25}0.32 & 0.69 \\
\textbf{Site C} & 0.56 & 0.33 & \cellcolor{gray!25}0.68 \\
\hline
\end{tabular}
\end{subtable}

\vspace{0.5em}

\begin{subtable}[t]{0.95\columnwidth}
\centering
\caption{Normalized BESS energy capacity,
$C_{\mathrm{batt}}/P_{\mathrm{PV}}^{\mathrm{rated}}$ (h).}
\label{tab:cross_dataset_energy}
\begin{tabular}{c|ccc}
\hline
\textbf{Train $\backslash$ Test}
& \textbf{Site A}
& \textbf{Site B}
& \textbf{Site C} \\
\hline
\textbf{Site A} & \cellcolor{gray!25}0.1 & 0.09 & 0.21 \\
\textbf{Site B} & 0.15 & \cellcolor{gray!25}0.07 & 0.3 \\
\textbf{Site C} & 0.1 & 0.07 & \cellcolor{gray!25}0.14 \\
\hline
\end{tabular}
\end{subtable}

\end{table}

To evaluate the generalization of CARL across different PV sites, the policy trained on each dataset was tested on all three site datasets. The BESS power and energy requirements were normalized by the rated PV capacity to enable comparison across plants of different sizes. The diagonal entries in Table~\ref{tab:cross_dataset_generalization} represent site-specific training and testing, while the off-diagonal entries indicate cross-dataset transfer. The results show that the peak BESS power requirement remains relatively consistent across different training datasets, indicating good generalization of the learned power-dispatch policy. The energy requirement is more sensitive to the training dataset, particularly when the policy is applied to Site~C, which exhibits the highest PV variability. Notably, the policy trained on Site~C transfers well to Sites~A and B, suggesting that training on data containing more diverse and severe PV fluctuations can improve generalization to less-variable operating conditions.

\subsection{Effect of BESS Efficiency on Energy Requirements}
\label{sec:efficiency_sizing}

\begin{figure}
  \centering
  \includegraphics[width=\columnwidth]{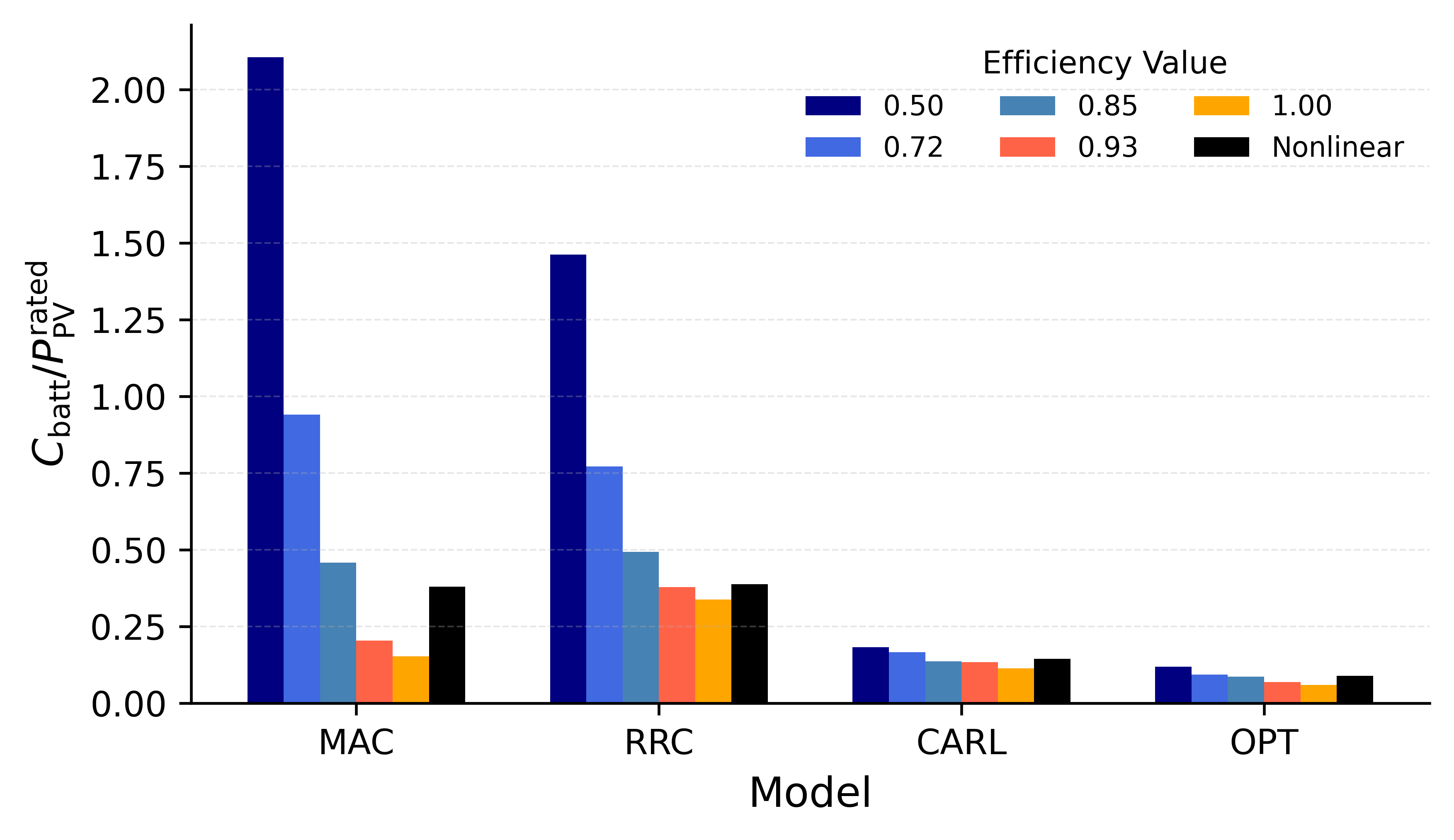}
  \caption{Required BESS energy capacity of the MAC, RRC, CARL, and ICNN
  controllers at Site~C for $\mathcal{R}_\mathrm{max}=10\,\%/\min$, obtained
  with constant efficiencies
  $\eta\in\{0.5,0.72,0.85,0.93,1\}$ and with the measured nonlinear
  efficiency--power characteristic of Section~\ref{sec:BESS model}.}
  \label{fig:energy_vs_eff}
\end{figure}

The sizing results presented in previous sections use the nonlinear efficiency--power characteristic described in Section~\ref{sec:BESS model}. To assess the influence of this modeling choice, the sizing procedure in Algorithm~\ref{alg:sizing} is repeated using the Site~C test set for $\mathcal{R}_{\max}=10\%/\min$ over a range of constant efficiencies, and the energy requirement results are shown in Fig.~\ref{fig:energy_vs_eff}. For all controllers, the required energy capacity increases as efficiency decreases. However, the magnitude of this increase depends on the smoothing strategy. 
MAC exhibits the greatest sensitivity to converter efficiency. Its normalized energy requirement increases from approximately $0.153$ at $\eta=1$ to $2.1$ at $\eta=0.5$. RRC increases substantially, from $0.338$ to $1.463$ over the same range. The stronger sensitivity of MAC is attributed to its frequent charging and discharging actions, which occur even when no ramp-rate violation is present. At high efficiency, MAC requires less energy capacity than RRC because its frequent charging and discharging actions can remain relatively balanced when conversion losses are negligible. However, as the efficiency decreases, these repeated actions accumulate larger conversion losses and increase the drift in the BESS SoC, causing the MAC energy requirement to grow more rapidly than that of RRC.

In contrast, CARL and ICNN are less sensitive to reductions in converter efficiency. This lower sensitivity is consistent with the explicit SoC-deviation terms in~\eqref{soc_reward} and~\eqref{opt_main_obj}, which encourage these controllers to regulate the SoC and limit the accumulation of loss-induced SoC deviations over the operating horizon. The results further show that a single constant-efficiency assumption does not reproduce the nonlinear-model sizing consistently across controllers. For example, at $\eta=0.93$, the constant-efficiency assumption underestimates the corresponding nonlinear-model energy requirements by approximately $46.4\%$, $2.4\%$, $7.8\%$, and $22.1\%$ for MAC, RRC, CARL, and ICNN, respectively.

\section{Conclusion}
\label{sec:conclusion}

This paper proposed a constraint-aware reinforcement learning (CARL) controller for reducing the BESS size required for PV ramp-rate smoothing. The proposed method constructs a causal feasibility interval from the ramp-rate, converter-power, and SoC constraints and maps the RL action into this interval, ensuring feasible operation by construction. A proximal policy optimization (PPO) agent is trained using a reward function that penalizes instantaneous battery power, SoC deviation, and increases in peak BESS power. A full-horizon input-convex neural network (ICNN)-based mixed-integer optimization problem provides a noncausal lower-bound benchmark for BESS sizing.

Evaluation using one-minute measurements from three PV sites showed that, at the $10\%/\mathrm{min}$ ramp-rate limit, CARL achieved the lowest BESS requirements among the evaluated causal controllers. Relative to conventional ramp-rate control, CARL reduced the estimated peak BESS power by 8.7--10.4\% and energy capacity by 47.5--62.6\% while maintaining ramp-rate compliance without requiring PV forecasts during deployment. These reductions translated into 17.0--22.4\% lower estimated BESS capital costs relative to conventional ramp-rate control under the adopted cost model. The results also highlight the importance of of reducing peak BESS power in PV smoothing applications, where power rating can dominate practical battery sizing and cost. The results further showed that converter efficiency can significantly affect the estimated energy requirement, supporting the use of power-dependent efficiency models in BESS sizing studies. Future work will incorporate additional objectives, including battery degradation and energy arbitrage, and validate the proposed controller through hardware-in-the-loop experiments.

\printcredits

\section*{Declaration of competing interest}

The authors declare that they have no known competing
financial interests or personal relationships that
could have appeared to influence the work reported
in this paper.

\section*{Acknowledgment}

This material is based upon work supported by the U.S. Department of Energy’s Office of EERE under award number DE-EE0010147. The views expressed herein do not necessarily represent the views of the U.S. Department of Energy or the United States Government.

\appendix
\section{Appendix: ICNN-Based Optimization Formulation}
\label{app:icnn_optimization}

This appendix presents the complete mathematical formulation of the noncausal ICNN-based optimization benchmark introduced in Section~\ref{sec:ICNN_opt}. For an ICNN with non-negative weights and convex nondecreasing ReLU activation functions, the network output is convex with respect to its input. The charging {\small($\eta_{\mathrm{ch}}[k] P_{\mathrm{batt}}^{\mathrm{ch}}[k] $) } and discharging {\small($- \frac{P_{\mathrm{batt}}^{\mathrm{dis}}[k]}{\eta_{\mathrm{dis}}[k]}$)} terms of~\eqref{Soc_model} are approximated using ICNNs~\cite{omidi2025towards}, denoted by $f_\mathrm{ICNN}(\cdot)$ and $g_\mathrm{ICNN}(\cdot)$, respectively. For a single-hidden-layer ICNN, the optimization formulation is represented by,

{\small
\begin{subequations}
\label{opt_full}
\begin{flalign}
&\underset{\substack{ P_\mathrm{batt}^\mathrm{ch}[k],\,P_\mathrm{batt}^\mathrm{dis}[k],\\
z^\mathrm{ch}[k],\,z^\mathrm{dis}[k]}}{\mathrm{minimize}}
\quad \left\|P_\mathrm{batt}^\mathrm{ch}
-P_\mathrm{batt}^\mathrm{dis}\right\|_{\infty}
+\lambda_{\mathrm{soc}}
\left\|\boldsymbol{E}_{\mathrm{batt}}
-E_{\mathrm{batt}}^{\mathrm{ref}}\right\|_{2}^{2}
&&\label{opt_full_obj}\\[2pt]
&\text{subject to}
&&\notag\\
&P_\mathrm{grid}[k] = P_\mathrm{PV}[k] -\big(P_\mathrm{batt}^\mathrm{ch}[k]-
P_\mathrm{batt}^\mathrm{dis}[k]\big)
&&\label{opt_full_grid}\\
&|P_\mathrm{grid}[k]- P_\mathrm{grid}[k-1]|\leq\Delta P_\mathrm{max}
&&\label{opt_full_ramp}\\[2pt]
&E_\mathrm{batt}[k+1]=E_\mathrm{batt}[k]+\tfrac{\Delta t}{60\times C_\mathrm{batt}}\big(
f_\mathrm{ICNN}[k]+g_\mathrm{ICNN}[k]\big)
&&\label{opt_full_soc}\\[2pt]
&f_\mathrm{ICNN}[k]=W_2^\mathrm{ch} z^\mathrm{ch}[k]+D_2^\mathrm{ch} P_\mathrm{batt}^\mathrm{ch}[k]+b_2^\mathrm{ch}
&&\label{opt_full_fch}\\
&z^\mathrm{ch}[k]=\text{max} (0, W_1^\mathrm{ch} P_\mathrm{batt}^\mathrm{ch}[k]+b_1^\mathrm{ch})
&&\label{opt_full_zch}\\
&g_\mathrm{ICNN}[k]=W_2^\mathrm{dis} z^\mathrm{dis}[k]+D_2^\mathrm{dis} P_\mathrm{batt}^\mathrm{dis}[k]+b_2^\mathrm{dis}
&&\label{opt_full_gdis}\\[2pt]
&z^\mathrm{dis}[k] = \text{max} (0, W_1^\mathrm{dis} P_\mathrm{batt}^\mathrm{dis}[k] + b_1^\mathrm{dis})
&&\label{opt_full_zdis}\\[2pt]
&\underline{E}_\mathrm{batt}\leq E_\mathrm{batt}[k] \leq \overline{E}_\mathrm{batt}
&&\label{opt_full_socbound}\\
&E_\mathrm{batt}[0]=E_\mathrm{batt}[N]=E_\mathrm{batt}^\mathrm{ref}
&&\label{opt_full_terminal}\\[2pt]
&0 \leq P_\mathrm{batt}^\mathrm{ch}[k],\, P_\mathrm{batt}^\mathrm{dis}[k] \leq P_\mathrm{batt}^\mathrm{max}&& \label{opt_full_powerbound}\\
&P_\mathrm{batt}^\mathrm{ch}[k] P_\mathrm{batt}^\mathrm{dis}[k] = 0
&&\label{opt_full_pcomp}
\end{flalign}
\end{subequations}}
where the objective in~\eqref{opt_full_obj} minimizes the peak BESS power while penalizing deviations of the SoC from its reference value, thereby targeting reductions in the required BESS power and energy capacities. The remaining constraints enforce the RRL~\eqref{opt_full_ramp}, BESS SoC dynamics~\eqref{opt_full_soc}-\eqref{opt_full_socbound}, SoC terminal condition~\eqref{opt_full_terminal}, and converter power limits~\eqref{opt_full_powerbound} and \eqref{opt_full_pcomp}. The terminal constraint in~\eqref{opt_full_terminal} ensures that the battery SoC returns to its reference value at the end of the optimization horizon. By using the complete daily PV profile, the optimization can proactively adjust the BESS SoC to satisfy this constraint.

The ICNN weights are trained using BESS efficiency data, as described in~\cite{omidi2025towards}. Although the ReLU function is convex, its equality constraint in~\eqref{opt_full_zch} does not define a convex constraint in the optimization formulation. Therefore, to embed the ICNN within the optimization problem, the ReLU relation in~\eqref{opt_full_zch} is represented using the Big-\(M\) mixed-integer formulation~\cite{omidi2025towards}. The~\eqref{opt_full_zch} can be represented by,\begin{subequations}
\begin{align}
    &z^\mathrm{ch}[k] \geq 0, \\
    &z^\mathrm{ch}[k] \geq W_1^\mathrm{ch} P_\mathrm{batt}^\mathrm{ch}[k] + b_1^\mathrm{ch}, \\
    &z^\mathrm{ch}[k] \leq M(1 - y^\mathrm{ch}[k]), \\
     &z^\mathrm{ch}[k] \leq W_1^\mathrm{ch} P_\mathrm{batt}^\mathrm{ch}[k] + b_1^\mathrm{ch}+ M y^\mathrm{ch}[k],
\end{align}
\label{opt_relax}
\end{subequations}
where the binary variable $y^\mathrm{ch}[k]\in\{0,1\}$ selects the active branch of the ReLU function. When $y^\mathrm{ch}[k]=0$, the constraints enforce $z^\mathrm{ch}[k]= W_1^\mathrm{ch} P_\mathrm{batt}^\mathrm{ch}[k] + b_1^\mathrm{ch}$, whereas $y^\mathrm{ch}[k]=1$ enforces $z^\mathrm{ch}[k]=0$. The constant $M$ is chosen sufficiently large to cover the feasible range of the affine term while avoiding unnecessarily large values that may adversely affect numerical performance~\cite{omidi2025towards}. The same formulation is applied to the discharging ICNN using $z^\mathrm{dis}[k]$ and $y^\mathrm{dis}[k]$. The nonconvex constraint in~\eqref{opt_full_pcomp}, which prevents simultaneous battery charging and discharging, is replaced by the following mixed-integer formulation:
\begin{align}
&0
\leq
P_\mathrm{batt}^\mathrm{ch}[k]
\leq
u[k]P_\mathrm{batt}^\mathrm{max}
&&\label{opt_full_pch}\\
&0
\leq
P_\mathrm{batt}^\mathrm{dis}[k]
\leq
(1-u[k])P_\mathrm{batt}^\mathrm{max},
&&\label{opt_full_pdis}
\end{align}
where $u[k]\in\{0,1\}$ is the binary variable which prevents simultaneous charging and discharging. Although the use of integer variables increases computational complexity, this is not a practical concern because the optimization is used only as an offline benchmark.

\bibliographystyle{cas-model2-names}

\bibliography{cas-refs}





\end{document}